\documentclass{article}
\usepackage{geometry}
\usepackage{wrapfig}
\usepackage{graphicx}
\usepackage{bm}
\usepackage{amsmath,amssymb}
\usepackage{mathrsfs}
\newcommand{\lw}[1]{\smash{\lower 1.5ex\hbox{#1}}}
\newcommand{\ri}[1]{\smash{\raise 1.5ex\hbox{#1}}}

\newcommand{\mapright}[1]{\smash{\mathop{\hbox to 1cm{\rightarrowfill}}\limits^{#1}}}
\newcommand{\diff}[2]{\frac{\partial #1}{\partial #2}}

\begin{document}

\title{Analyses of multiplicity distributions and Bose--Einstein correlations at the LHC using negative binomial distribution and generalized Glauber--Lachs formula}

\author{Minoru Biyajima$^{1}$, and Takuya Mizoguchi$^{2}$\\
{\small $^{1}$Department of Physics, Shinshu University, Matsumoto 390-8621, Japan}\\
{\small $^{2}$National Institute of Technology, Toba College, Toba 517-8501, Japan}}

\maketitle

\begin{abstract}
This study aims to analyze the data on multiplicity distributions and Bose--Einstein correlations (BEC) collected at the LHC by the ATLAS and CMS Collaborations using a double-generalized Glauber--Lachs formula (D-GGL) and double-negative binomial distribution (D-NBD). From this investigation, it can be inferred that the D-GGL formula performs as effectively as the D-NBD. Moreover, our results show that the parameters estimated in multiplicity distributions (MD) ($P(n)$) are related to those contained in the BEC formula.
\end{abstract}


\section{\label{sec1}Introduction}
Recently various kinds of data on multiplicity distributions (MD ($P(n)$)) with pseudorapidity intervals ($|\eta|<\eta_c$) at LHC energies have been reported \cite{Aad:2010ac,Aaboud:2016itf,Aamodt:2010pp,Khachatryan:2010nk}. The double-negative binomial distribution (D-NBD) formula has been utilized to analyze these data \cite{Ghosh:2012xh,Zaccolo:2015udc}. The D-NBD formula was originally proposed in \cite{Fuglesang:1989st} to explain the violation of KNO scaling \cite{Koba:1972ng} at $\sqrt s = 900$ GeV observed by the UA5 collaboration \cite{Ansorge:1988kn}. The D-NBD is expressed as follows:
\begin{eqnarray}
  P(n,\, \langle n\rangle) = \alpha P_{\rm NBD_1}(n,\,k_{\rm N_1},\,\langle n_1\rangle) + (1-\alpha)P_{\rm NBD_2}(n,\,k_{\rm N_2},\,\langle n_2\rangle),
\label{eq1}
\end{eqnarray}
where $\alpha$ is the weight factor for the first NBD \cite{Giovannini:1998zb,Giovannini:2003ft,Dremin:2004ts,Zborovsky:2013tla}. The notations, $k_1$ and $k_2$, represent intrinsic parameters contained in the NBDs, and $\langle n_1\rangle$ and $\langle n_2\rangle$ are the averaged multiplicities. The NBD is given by the following equation:
\begin{eqnarray}
 P_{\rm NBD}(n,\,k_{\rm N},\,\langle n\rangle) = \frac{\Gamma (n+k_{\rm N})}{\Gamma (n+1)\Gamma (k_{\rm N})}\frac{(\langle n\rangle/k_{\rm N})^n}{(1+\langle n\rangle/k_{\rm N})^{n+k_{\rm N}}},
\label{eq2}
\end{eqnarray}

On the other hand, in \cite{Mizoguchi:2010vc} we analyzed various data with $|\eta|\le \eta_c$ on MD ($P(n)$) in terms of the NBD and generalized Glauber--Lachs (GGL) formula \cite{Biyajima:1982un,Biyajima:1984aq}. In that work, we found that the role of the GGL formula is compatible with that of the NBD. The GGL formula is given as follows,
\begin{eqnarray}
P_{\rm GGL}(n,\,k_{\rm G},\,p,\,\langle n\rangle) = \frac{(p\langle n\rangle/k_{\rm G})^n}{(1+p\langle n\rangle/k_{\rm G})^{n+k_{\rm G}}}
\exp\left[-\frac{\gamma p\langle n\rangle}{1+p\langle n\rangle/k_{\rm G}}\right]
L_n^{(k_{\rm G}-1)}\left(-\frac{\gamma k_{\rm G}}{1+p\langle n\rangle/k_{\rm G}}\right),
\label{eq3}
\end{eqnarray}
where $p=1/(1 + \gamma)$ with the ratio of $\gamma= \langle n_{\rm coherent}\rangle/\langle n_{\rm chaotic}\rangle$. It should be noted that the average coherent and chaotic multiplicities are contained in Eq. (\ref{eq3}). Here, $k_{\rm G}$ is also an intrinsic parameter of the GGL formula. The analyses in \cite{Mizoguchi:2010vc} suggested that a finite $\gamma$ indicates that the coherent component seems to be necessary in data with $|\eta|<\eta_c$. We remark that Eq. (\ref{eq3}) has the following stochastic property.
\begin{eqnarray}
\mbox{Eq. (\ref{eq3})}\ \left\{
\begin{array}{l}
\mapright{\quad\ k_{\rm G}=1\qquad}\ \mbox{original GL}\ \mapright{\quad\ p\to 1\qquad}\ \mbox{Furry dis.}\medskip\\
\mapright{\ k_{\rm G}=k_{\rm N},\ p\to 1\ }\ \mbox{NBD of Eq. (\ref{eq2})}\\
\hspace{30mm} \downarrow k_{\rm N}\to \infty\\
\mapright{\quad\ p\to 0\qquad}\ \mbox{Poisson distribution}
\end{array}
\right.
\label{eq4}
\end{eqnarray}
In the appendix \ref{secA}, the stochastic background of Eqs. (\ref{eq2}) and (\ref{eq3}) is presented. 

Thus, in this paper, we propose the following double-GGL formula (D-GGL) to analyze data on charged MD ($P(n)$) at LHC energies:
\begin{eqnarray}
  P(n,\, \langle n\rangle) = \alpha P_{\rm GGL_1}(n,\,k_{\rm G}=2,\,p_1,\,\langle n_1\rangle) + (1-\alpha)P_{\rm GGL_2}(n,\,k_{\rm G}=2,\,p_2,\,\langle n_2\rangle),
\label{eq5}
\end{eqnarray}
where $p_i$, $\langle n_i\rangle$ ($i=1,\ 2$) are introduced to distinguish the parameters contained in D-GGL formula. It should be noted that $k_{\rm G}=2$ reflects the degree of freedom for the $(+-)$ charged particle ensembles.


\begin{figure}[htbp]
  \centering
  \includegraphics[width=0.35\columnwidth]{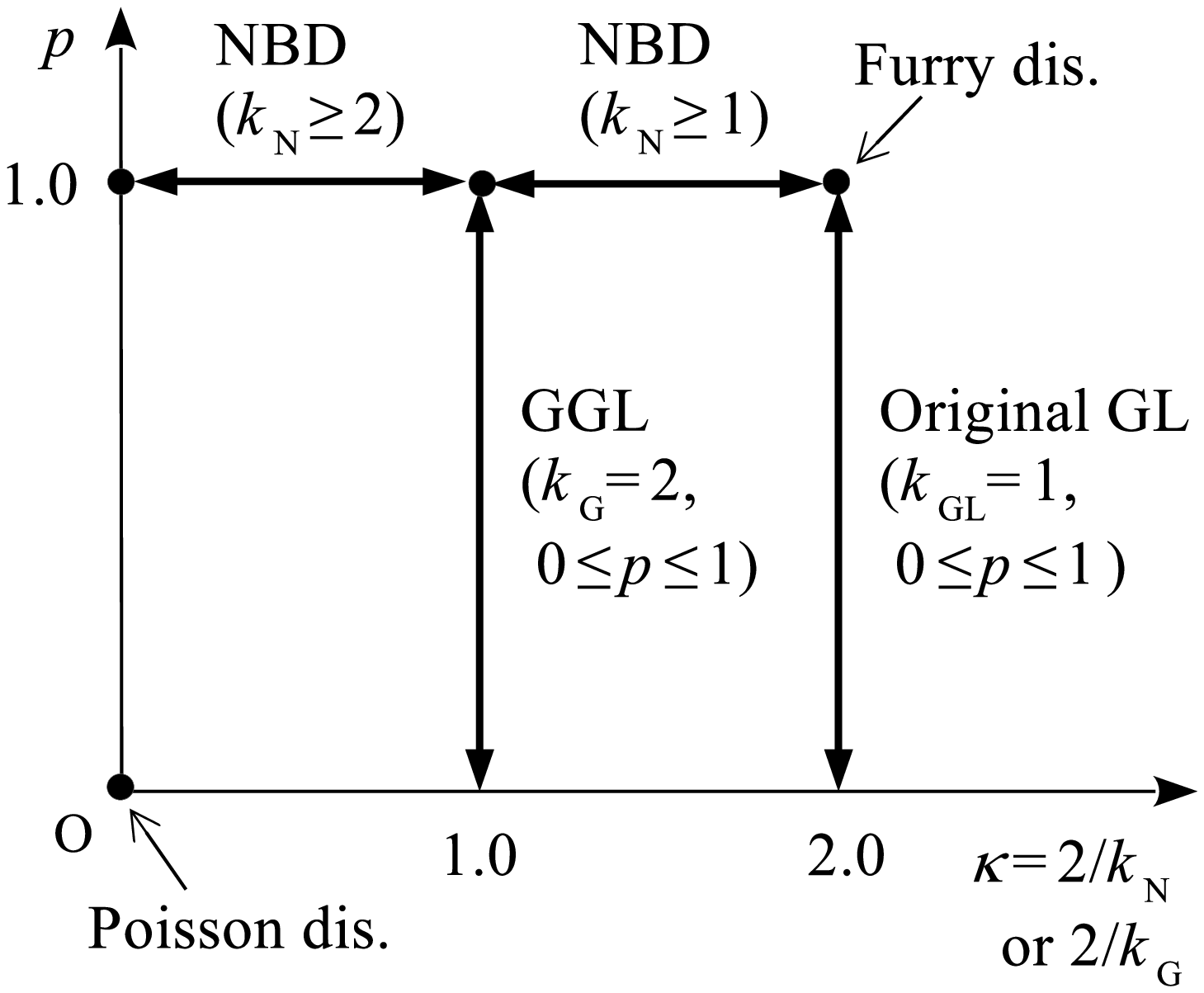}\qquad\qquad
  \includegraphics[width=0.40\columnwidth]{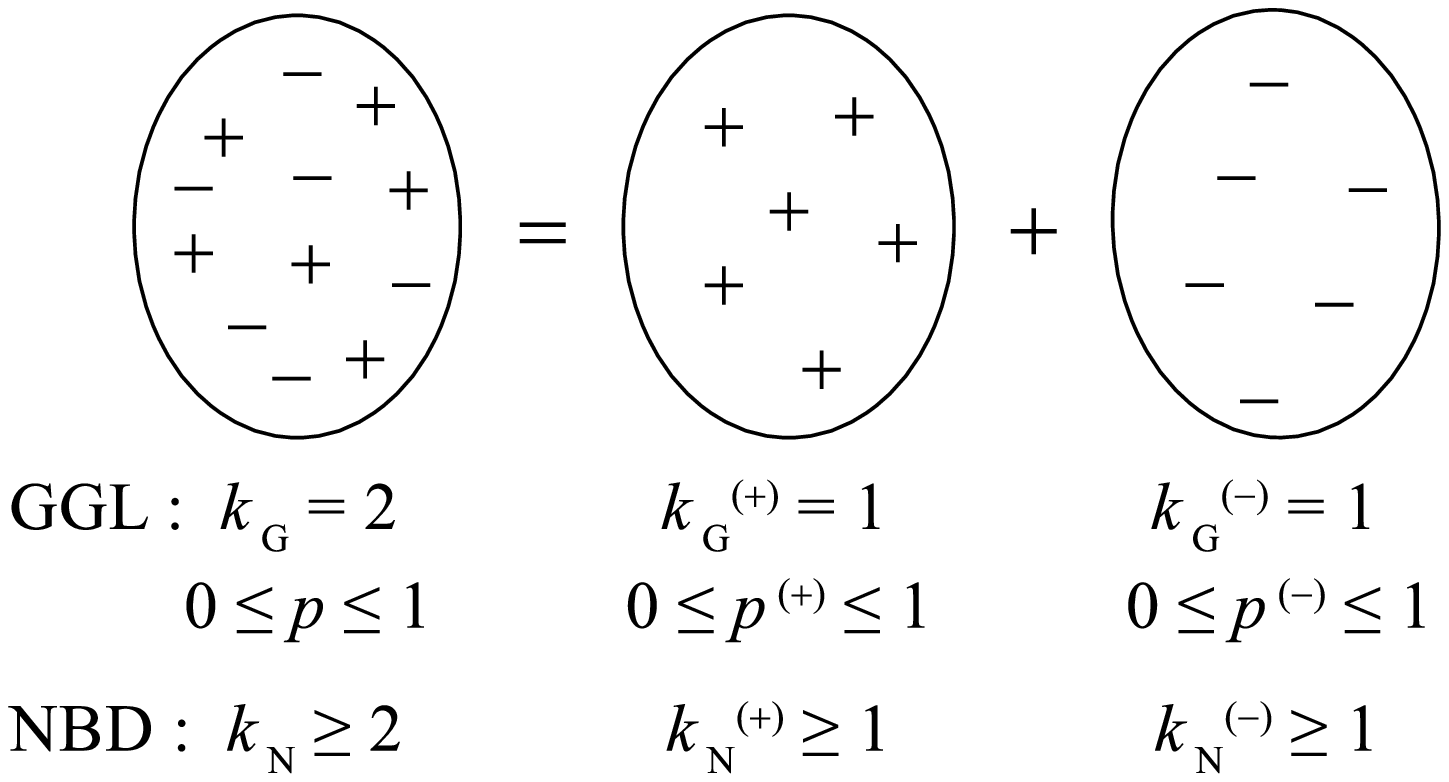}\\
  (a)\hspace{70mm} (b)
  \caption{\label{fig1}(a) The stochastic properties of the generalized Glauber--Lachs formula are shown in the $\kappa$-$p$ plane. At the point (2.0, 1.0), the perfect Bose--Einstein statistics for the same charged particles ensemble holds. At the point (1.0, 1.0), the Bose--Einstein statistics for the $(+-)$ charged particle ensemble holds. (b) The charged particle ensemble is decomposed into the positive $(+)$ and negative $(-)$ ensembles. Here $\gamma= \langle n_{\rm coherent}\rangle/\langle n_{\rm chaotic}\rangle$ can be assumed to reflect the contamination ($K$, $p$ / $\pi$'s), and the degree of superposition of the phase spaces of particles. }
\end{figure}

As a next step, we would like to consider the Bose--Einstein correlations (BECs). The BEC for positive charged particles \cite{Biyajima:1990ku} is given as follows:
\begin{eqnarray}
 N_{\rm GL}^{(2+)}/N^{\rm BG} = \left\{1 + \frac 1{k_{\rm GL}^{(+)}}[2p^{(+)}(1-p^{(+)})E_{\rm BE}^{1/2} + (p^{(+)})^2E_{\rm BE}]\right\},
\label{eq6}
\end{eqnarray}
where the left-hand side represents the normalized number of pairs of positive charged particles, and $E_{\rm BE}$ is the exchange function between them,
\begin{eqnarray}
E_{\rm BE} =\ \left\{
\begin{array}{l}
\exp(-RQ),\medskip\\
\exp(-(RQ)^2),
\end{array}
\right.
\label{eq7}
\end{eqnarray}
where $Q=\sqrt{-(p_1-p_2)^2}$, and $R$ denotes the interaction range between two particles. Hereafter, for the concrete analysis of the BEC we employ the exponential form, because values of $\chi^2$'s for the Gaussian formula are larger than those of the exponential formula.

As seen in Fig. \ref{fig1}, $k_{\rm GL}^{(+)} = k_{\rm GL}^{(-)} = 1$ and $p = p^{(+)} = p^{(-)}$ are utilized. For the D-GGL formula, we can calculate the following formula for the BEC with weight factor $\alpha$, provided that MD ($P_{\rm GGL_1}(n)$) and ($P_{\rm GGL_2}(n)$) are independent ensembles, 
\begin{eqnarray}
 N^{(2+:\,2-)}_{\rm D-GGL}/N^{\rm BG} = \left\{1 + \alpha[2p_1(1-p_1)E_{\rm BE_1}^{1/2} + p_1^2E_{\rm BE_1}] + (1-\alpha)[2p_2(1-p_2)E_{\rm BE_2}^{1/2} + p_2^2E_{\rm BE_2}]\right\}.
\label{eq8}
\end{eqnarray}
In Eq. (\ref{eq8}), since $p_1=p_2=1.0$ and using the replacements $k_{\rm N_1}^{(+)} = k_{\rm N_1}^{(-)} = 2/k_{\rm N_1}$ and $k_{\rm N_2}^{(+)} = k_{\rm N_2}^{(-)} = 2/k_{\rm N_2}$, we obtain the following BEC for the D-NBD:
\begin{eqnarray}
 N^{(2+:\,2-)}_{\rm D-NBD}/N^{\rm BG} = \left\{1 + \alpha (2/k_{\rm N_1})E_{\rm BE_1} + (1-\alpha)(2/k_{\rm N_2})E_{\rm BE_2}\right\}.
\label{eq9}
\end{eqnarray}
In the appendix \ref{secB}, Eq. (\ref{eq9}) is derived from a stochastic approach. In Eq. (\ref{eq9}), when $E_{\rm BE_1}$ and $E_{\rm BE_2}$ are the same functions, we obtain the conventional formula (CF)
\begin{eqnarray}
 N^{(2+:\,2-)}_{\rm CF}/N^{\rm BG} = \left\{1 + \lambda_{\rm eff}E_{\rm BE}\right\}.
\label{eq10}
\end{eqnarray}
provided that the coefficient of $\lambda_{\rm eff} = 2[\alpha/k_{\rm N_1} + (1-\alpha)/k_{\rm N_2}]$ is assumed to be a free parameter.

By employing Eqs. (\ref{eq1})$\sim$(\ref{eq5}) for the MD ($P(n)$), it can be expected that we can analyze the data on MD ($P(n)$) and the BEC concurrently.

The present paper is organized as follows. In sect. 2, data on the MD ($P(n)$) at the LHC collected by the ATLAS collaboration are treated. In sect. 3, data on the BEC collected by the ATLAS collaboration are performed. In the fourth section, data on the MD ($P(n)$) and the BEC by the CMS collaboration are analyzed. In the final section, concluding remarks and discussions are presented. In the Appendix \ref{secA}, the stochastic background of the NBD and the GGL formula is shown. In the Appendix \ref{secB}, the derivation of the BEC in the two-component model is presented.


\section{\label{sec2}Analyses of MD at $\sqrt s =$0.9, 7, 8, and 13 TeV by ATLAS data}
\begin{figure}[htbp]
  \centering
  \includegraphics[width=0.48\columnwidth]{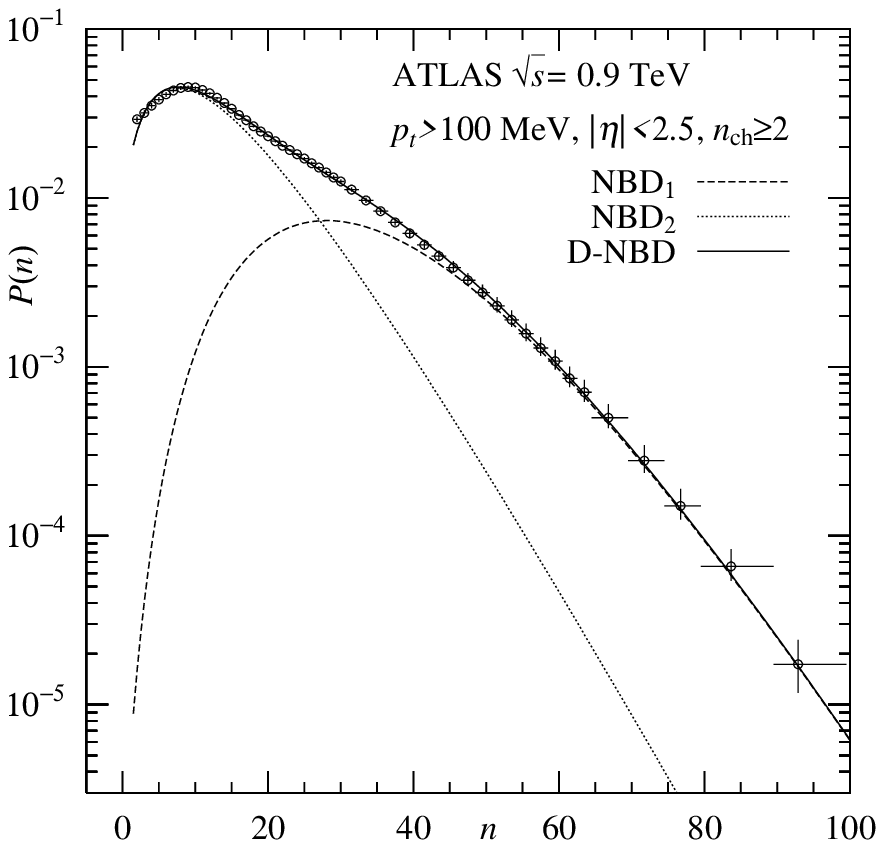}
  \includegraphics[width=0.48\columnwidth]{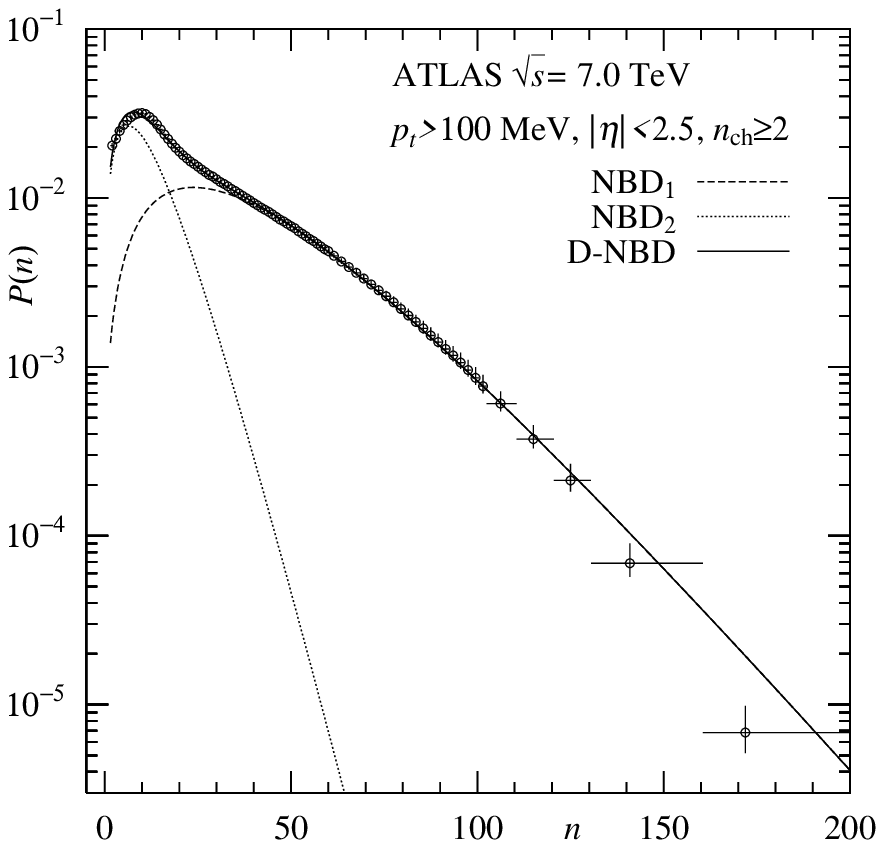}\\
  \includegraphics[width=0.48\columnwidth]{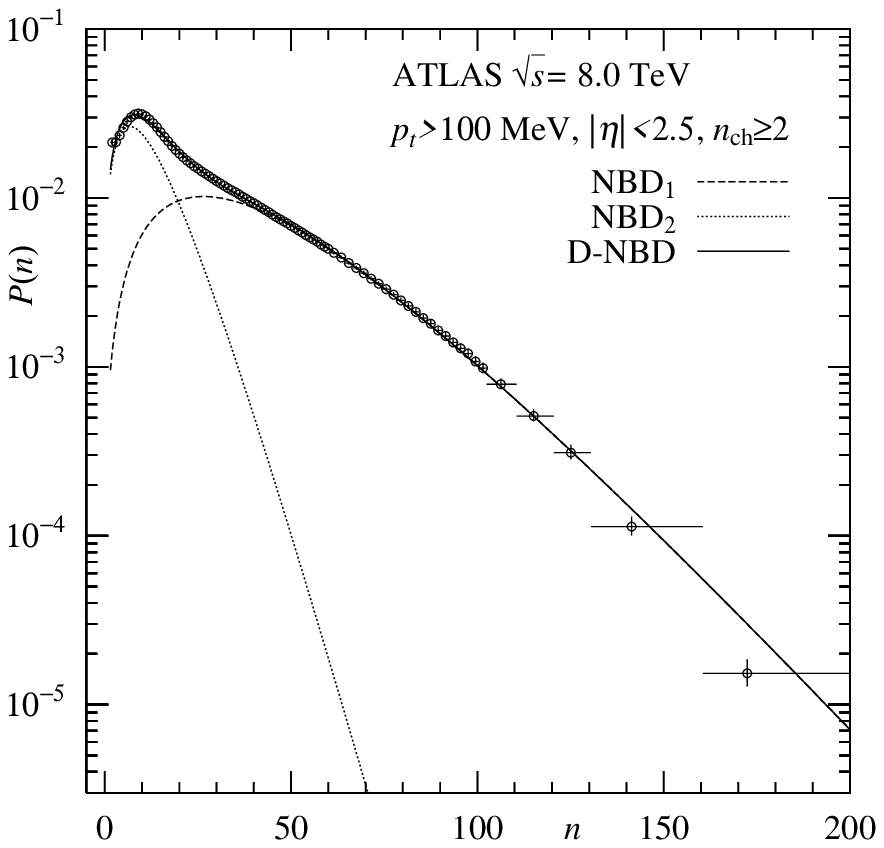}
  \includegraphics[width=0.48\columnwidth]{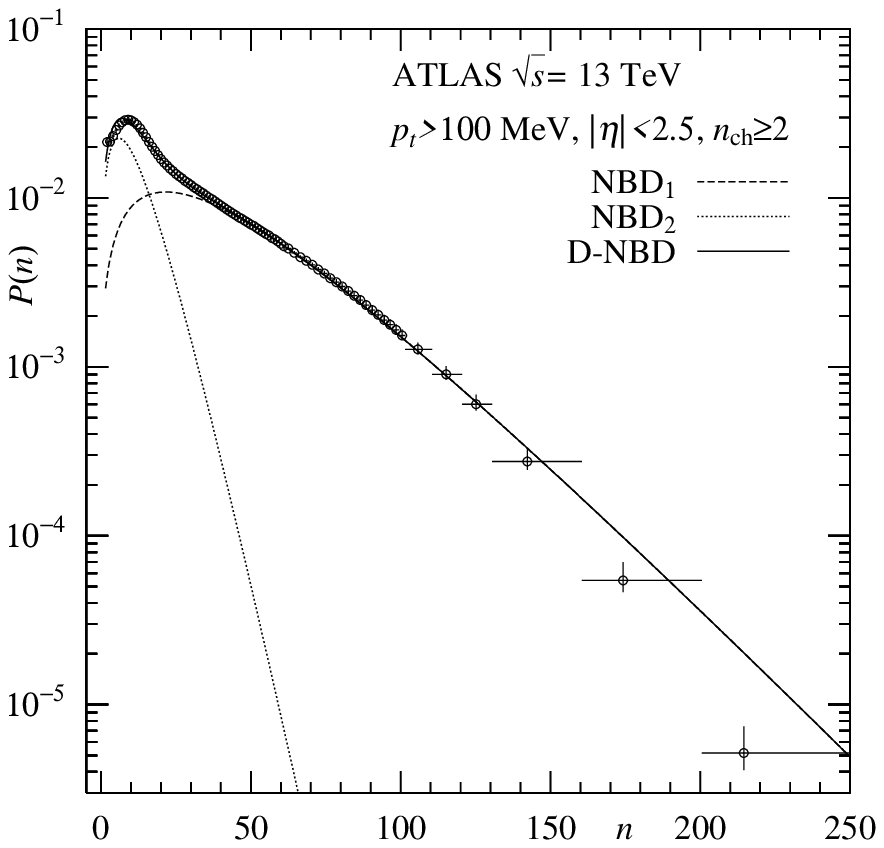}
  \caption{\label{fig2}Analyses of ATLAS data ($p_t>100$ MeV, $|\eta| < 2.5$, $n_{\rm ch} \ge 2$) using Eq. (\ref{eq1}).}
\end{figure}

\begin{figure}[htbp]
  \centering
  \includegraphics[width=0.48\columnwidth]{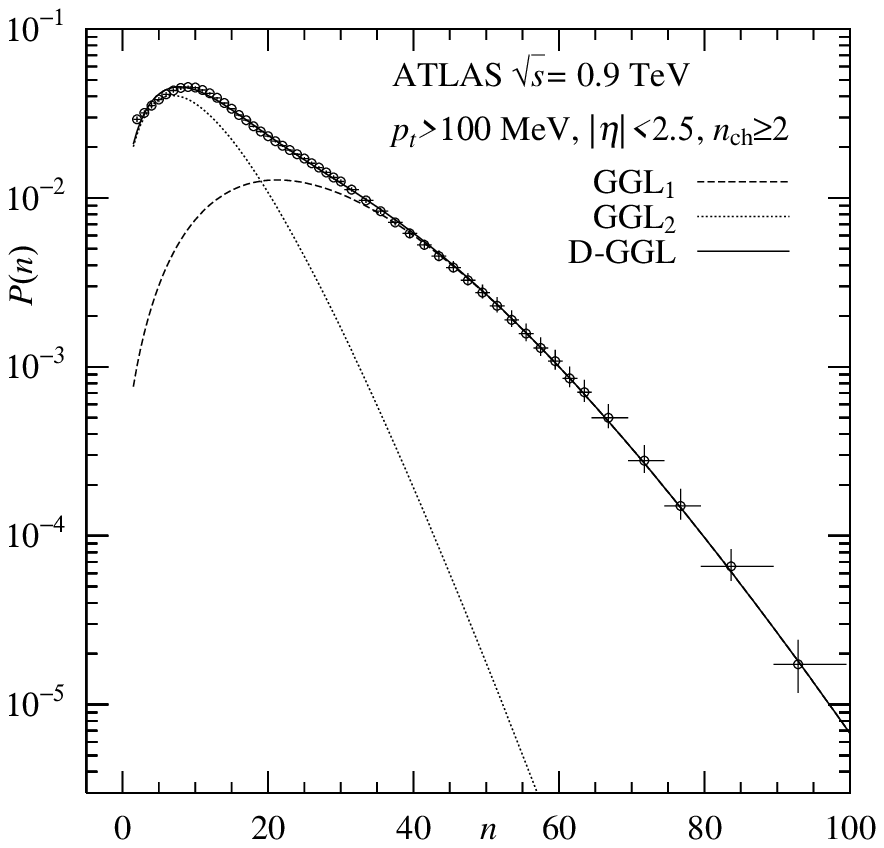}
  \includegraphics[width=0.48\columnwidth]{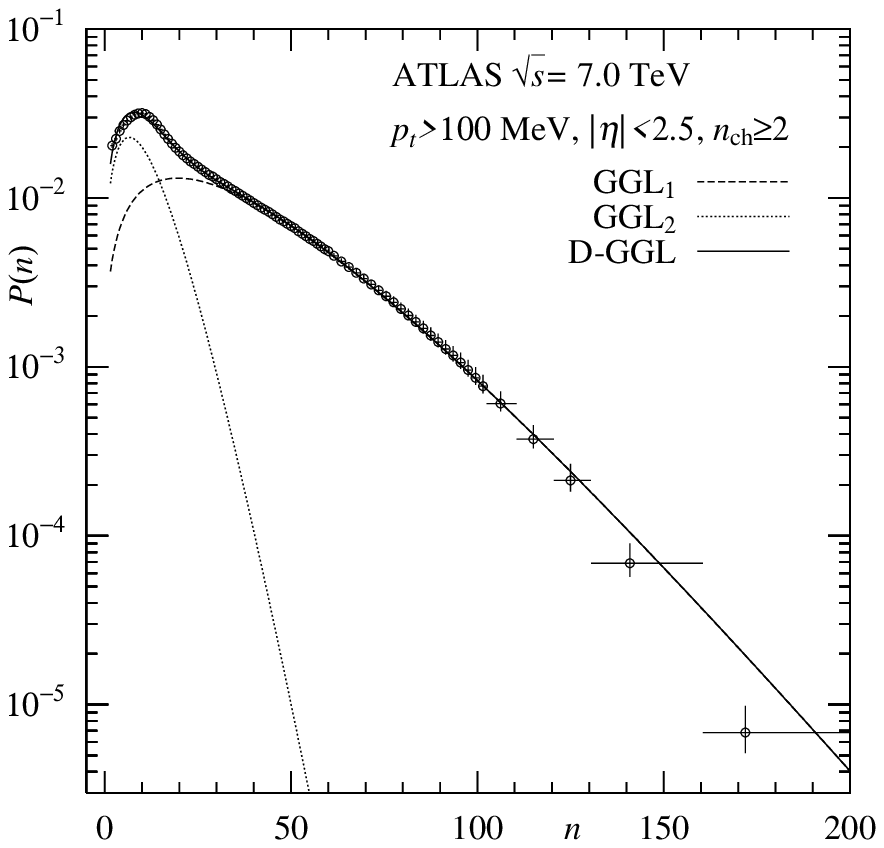}\\
  \includegraphics[width=0.48\columnwidth]{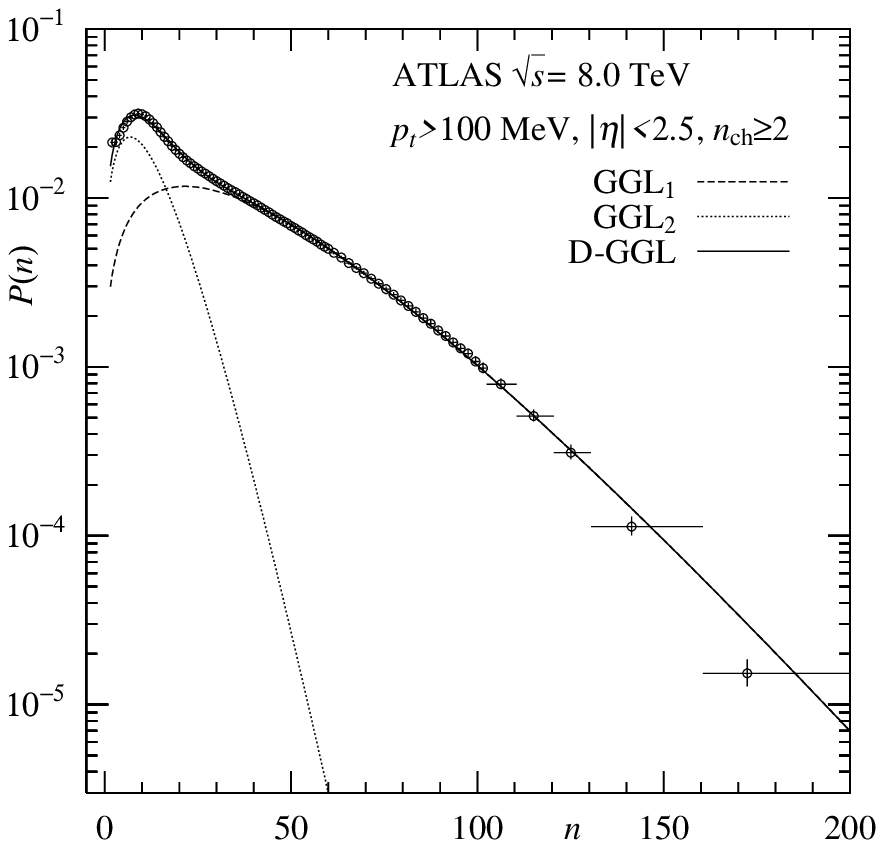}
  \includegraphics[width=0.48\columnwidth]{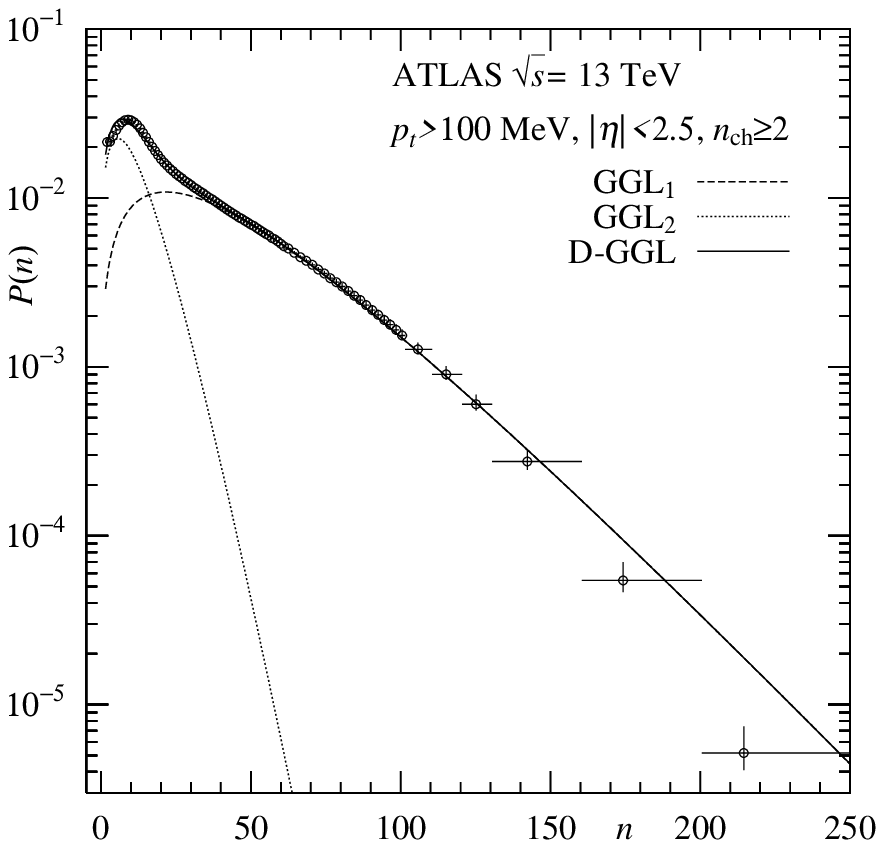}
  \caption{\label{fig3}Analyses of ATLAS data ($p_t>100$ MeV, $|\eta| < 2.5$, $n_{\rm ch} \ge 2$) using Eq. (\ref{eq5}).}
\end{figure}

By employing Eqs. (\ref{eq1}) and (\ref{eq5}), we can analyze data on MD collected by the ATLAS collaboration~\cite{Aad:2010ac} at LHC energies.  Our results are presented in Figs. \ref{fig2} and \ref{fig3} and Table \ref{tab1}.


\begin{table}[htbp]
\centering
\caption{\label{tab1}Analyses of MD ($P(n)$) collected by the ATLAS collaboration ($p_t>100$ MeV, $|\eta| < 2.5$, $n_{\rm ch} \ge 2$) using Eqs. (\ref{eq1}) and (\ref{eq5}).} 
\vspace{1mm}
\begin{tabular}{c|cccccc}
\hline
MD ($P(n)$ & \multicolumn{6}{c}{Eq. (\ref{eq1})\quad D-NBD}\\
$\sqrt s$ [TeV] & $\alpha$ & $k_1$ & $\langle n_1\rangle$ & $k_2$ & $\langle n_2\rangle$ & $\chi^2/$n.d.f.\\
\hline
0.9
& 0.23$\pm$0.03
& 7.52$\pm$0.77
& 32.9$\pm$1.3 
& 2.70$\pm$0.12
& 12.6$\pm$0.4
& 70.2/46\\
7.0
& 0.60$\pm$0.02
& 2.61$\pm$0.10
& 39.4$\pm$0.6
& 2.72$\pm$0.09
& 11.0$\pm$0.1
& 129/80\\
8.0
& 0.57$\pm$0.02
& 2.70$\pm$0.12
& 42.7$\pm$0.8
& 2.58$\pm$0.10
& 11.9$\pm$0.2
& 119/80\\
13
& 0.66$\pm$0.00
& 2.00$\pm$0.02
& 44.2$\pm$0.2
& 2.46$\pm$0.06
& 10.9$\pm$0.1
& 263/81\\
\hline
& \multicolumn{6}{c}{Eq. (\ref{eq5})\quad D-GGL}\\
$\sqrt s$ [TeV] & $\alpha$ & $p_1$ & $\langle n_1\rangle$ & $p_2$ & $\langle n_2\rangle$ & $\chi^2/$n.d.f.\\
\hline
0.9
& 0.41$\pm$0.05
& 0.25$\pm$0.04
& 27.0$\pm$1.4
& 0.39$\pm$0.02
& 10.6$\pm$0.4
& 60.3/46\\
7.0
& 0.67$\pm$0.01
& 0.68$\pm$0.03
& 36.7$\pm$0.3
& 0.42$\pm$0.02
& 10.4$\pm$0.1
& 114/80\\
8.0
& 0.65$\pm$0.01
& 0.66$\pm$0.03
& 39.5$\pm$0.4
& 0.45$\pm$0.03
& 11.1$\pm$0.2
& 111/80\\
13
& 0.66$\pm$0.00
& 0.90$\pm$0.06
& 44.1$\pm$0.2
& 0.60$\pm$0.03
& 10.7$\pm$0.1
& 280/81\\
\hline
\end{tabular}
\end{table}


\section{\label{sec3}Analysis of Bose--Einstein correlation at $\sqrt s =$ 0.9 and 7.0 TeV for ATLAS data} 
\begin{figure}[htbp]
  \centering
  \includegraphics[width=0.48\columnwidth]{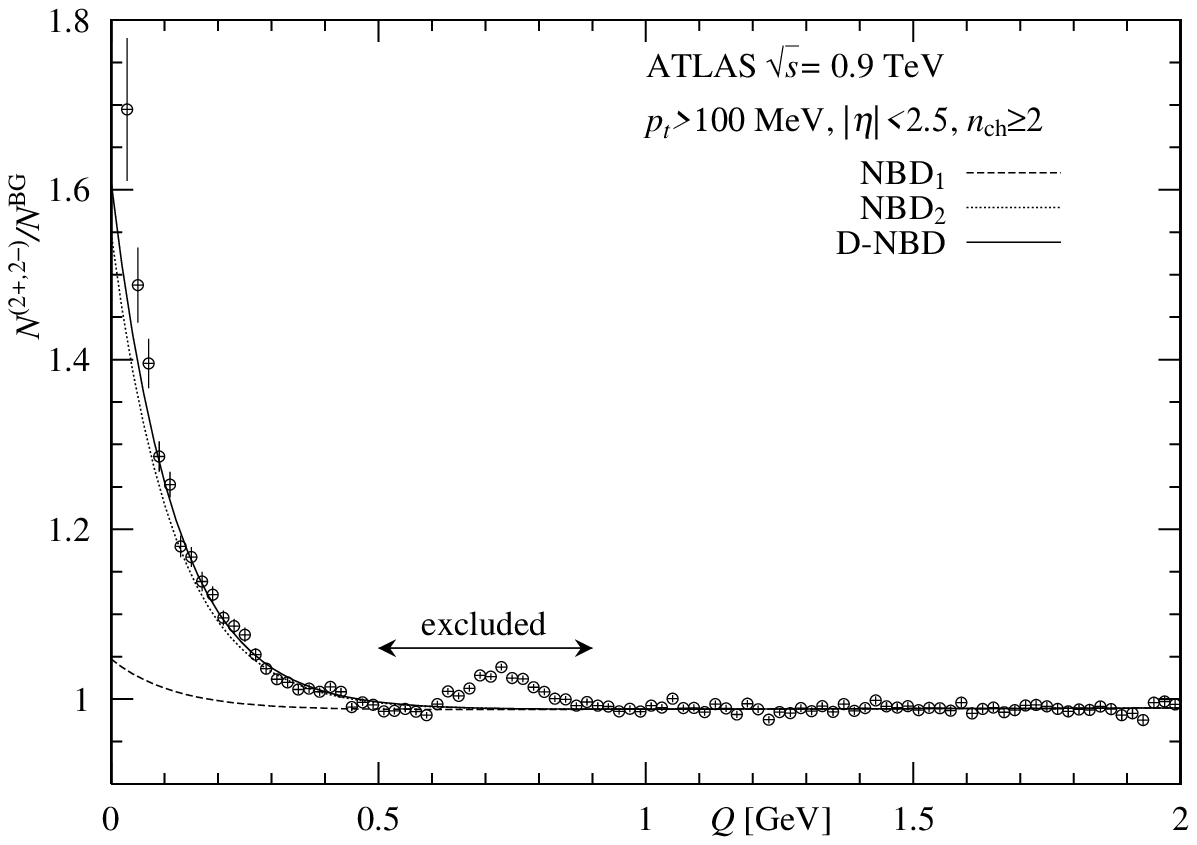}
  \includegraphics[width=0.48\columnwidth]{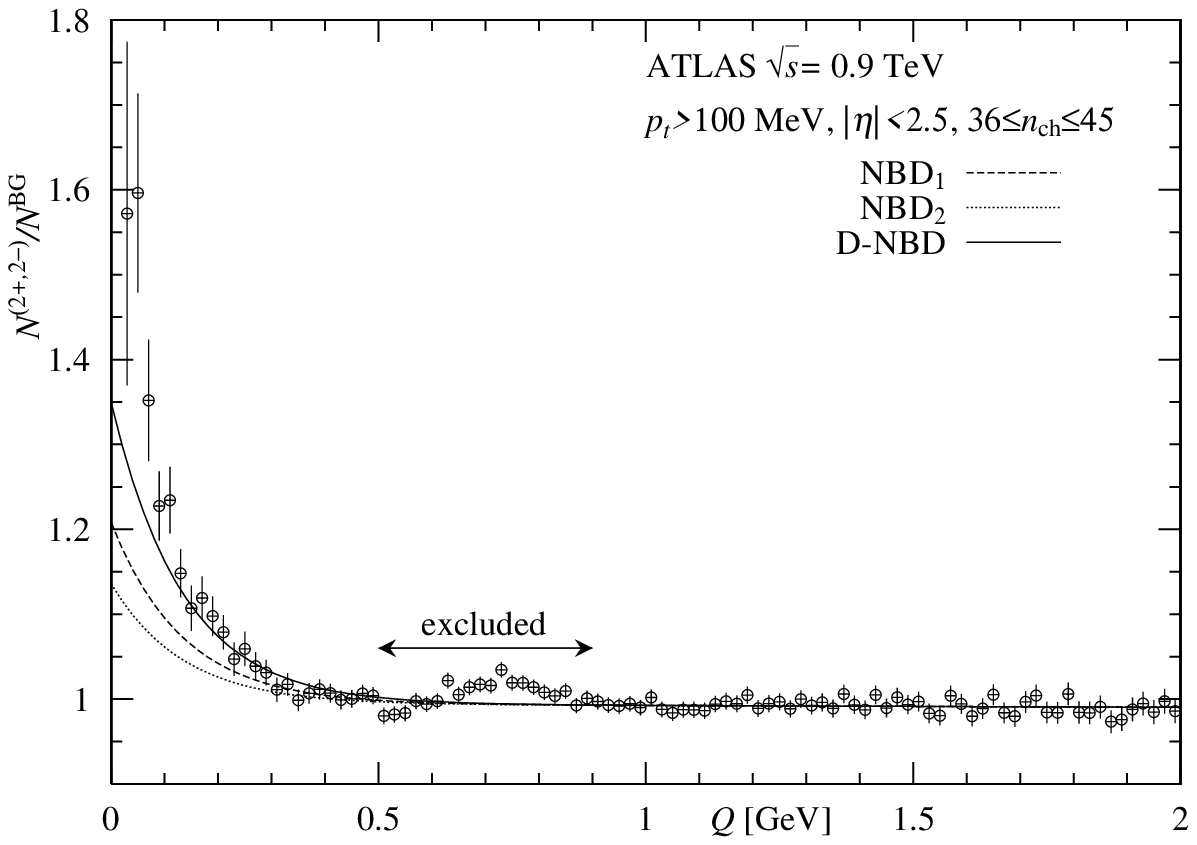}\\
  \includegraphics[width=0.48\columnwidth]{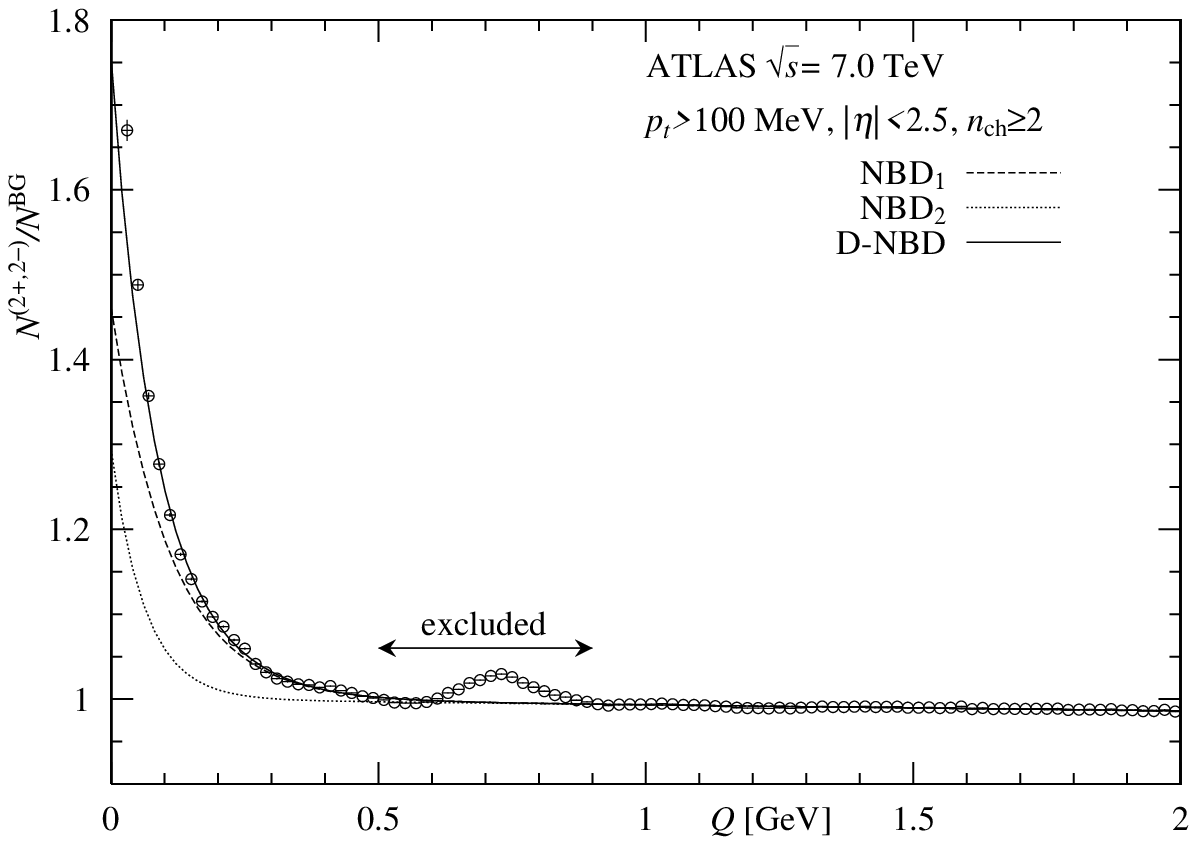}
  \includegraphics[width=0.48\columnwidth]{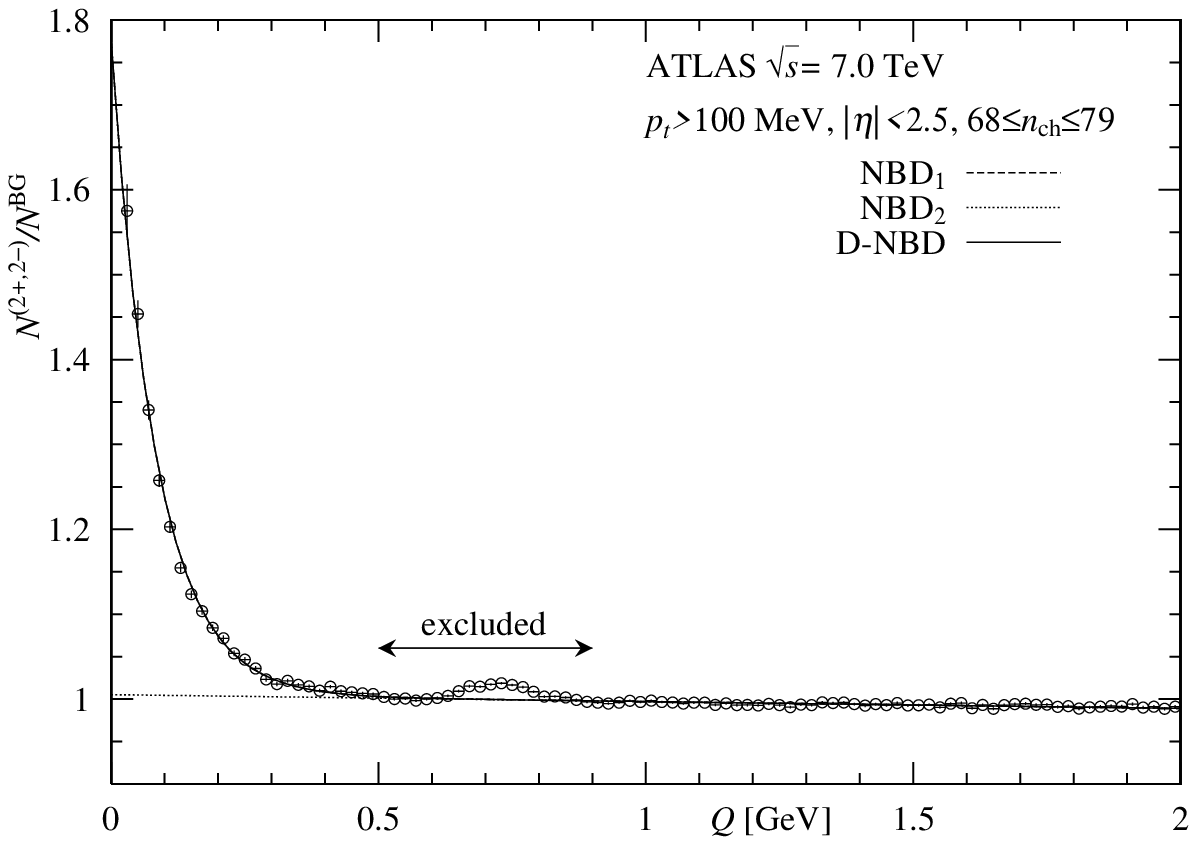}\\
  \includegraphics[width=0.48\columnwidth]{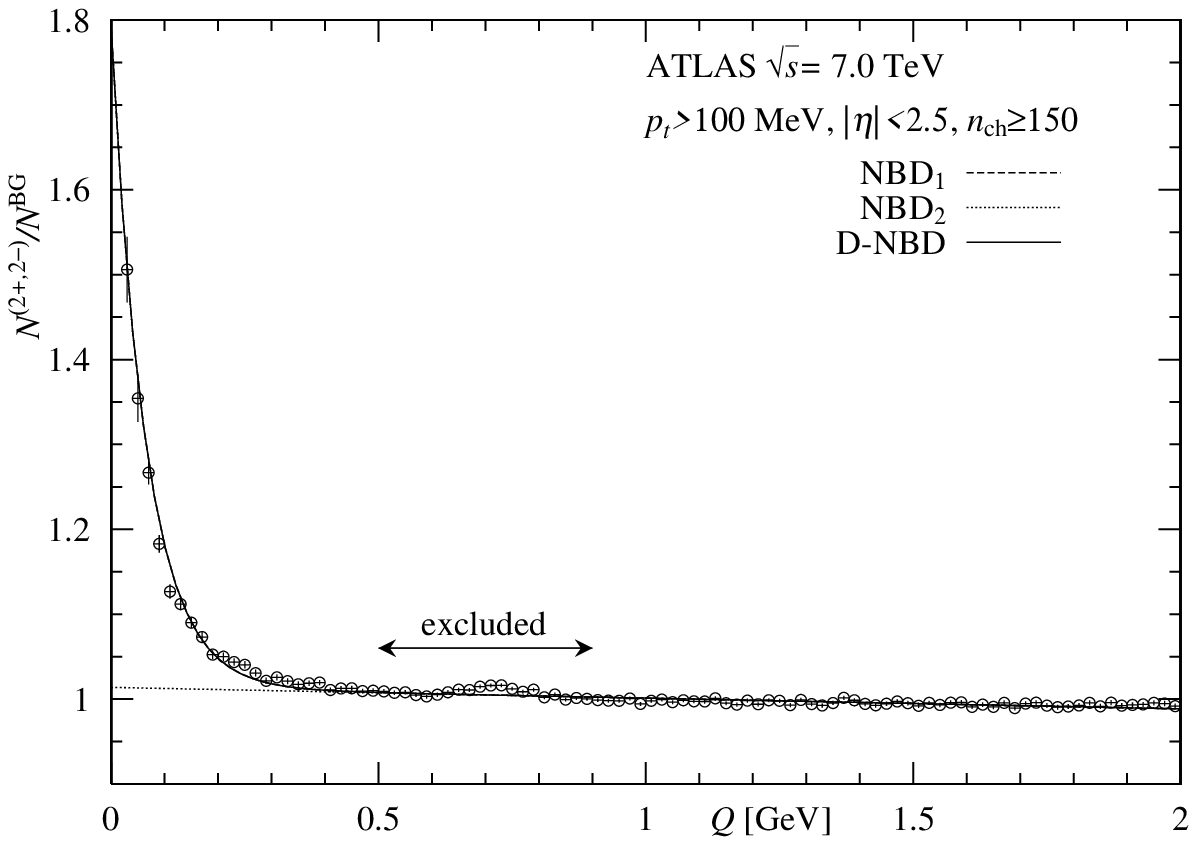}
  \caption{\label{fig4}Analyses of ATLAS data on the BEC ($p_t>100$ MeV, $|\eta| < 2.5$, $n_{\rm ch} \ge 2$) using Eq. (\ref{eq9}) with $E_{\rm BE} = \exp(-RQ)$.}
\end{figure}

\begin{figure}[htbp]
  \centering
  \includegraphics[width=0.48\columnwidth]{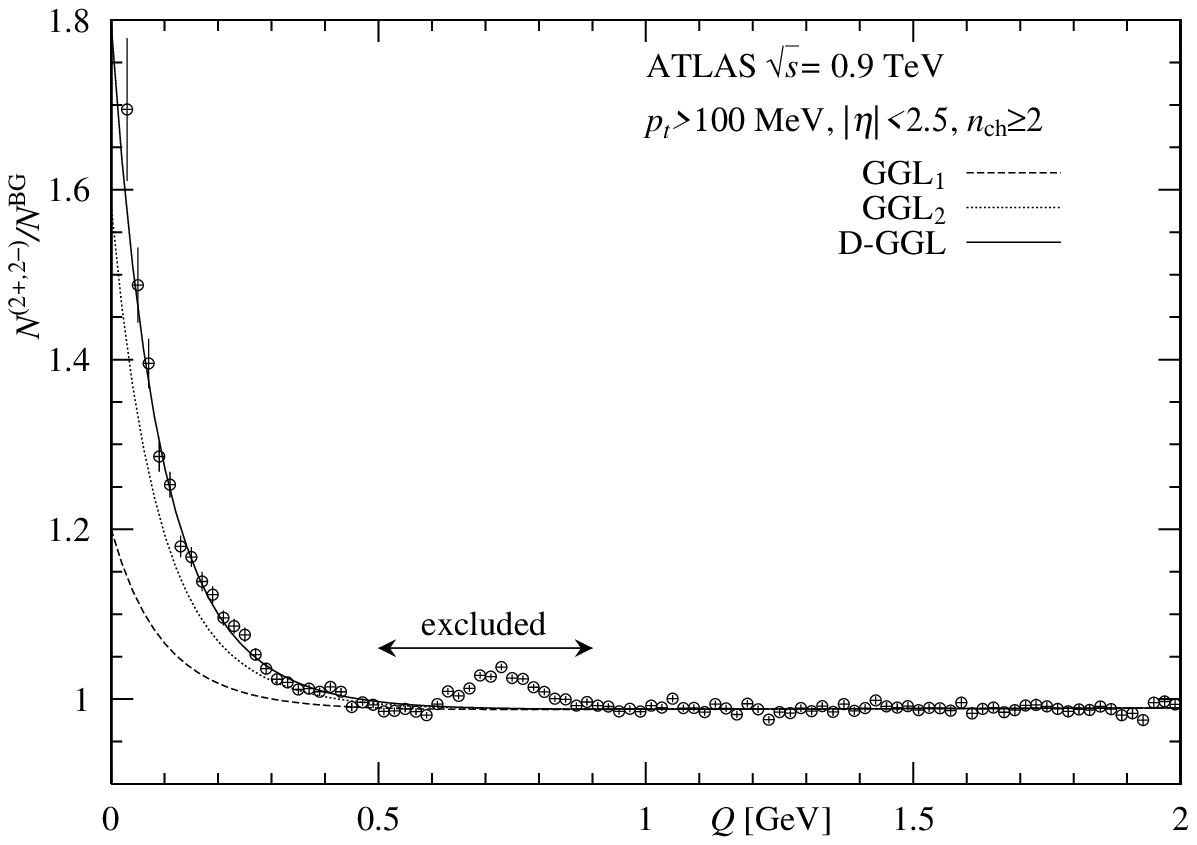}
  \includegraphics[width=0.48\columnwidth]{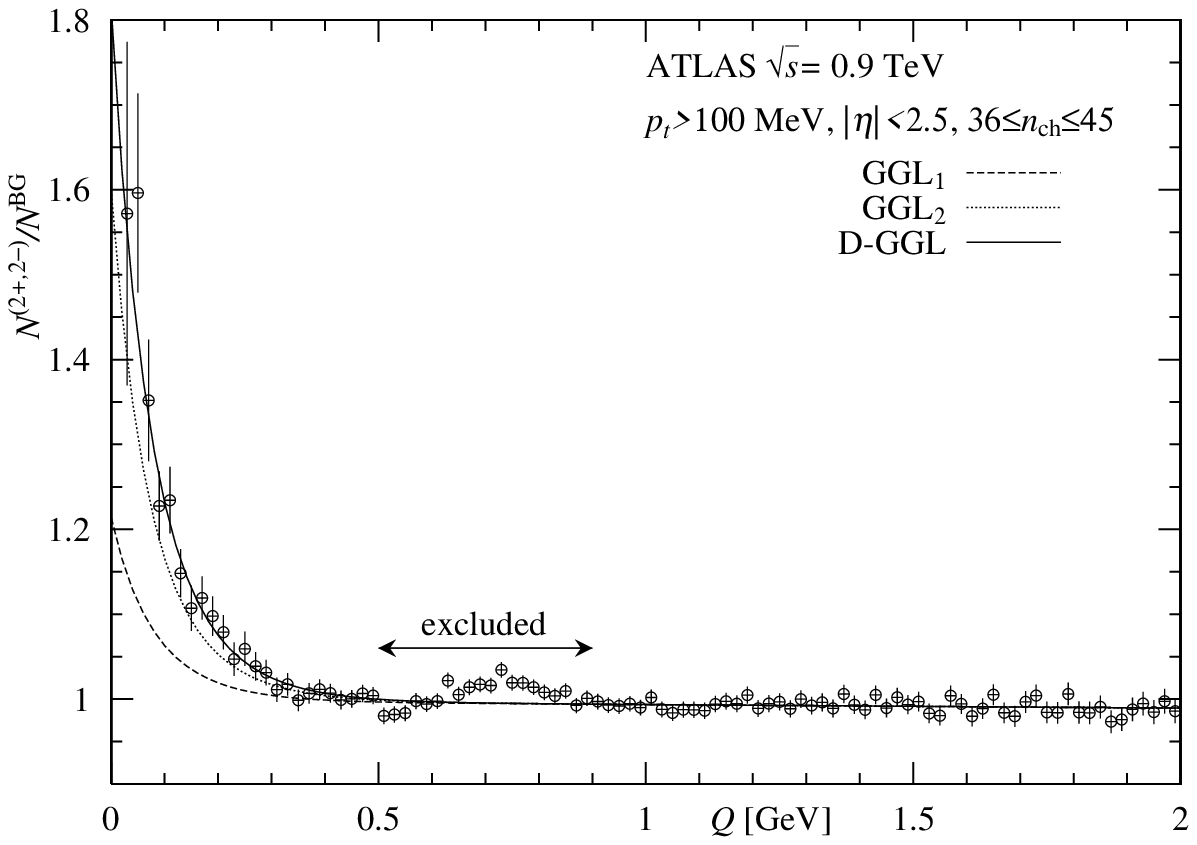}\\
  \includegraphics[width=0.48\columnwidth]{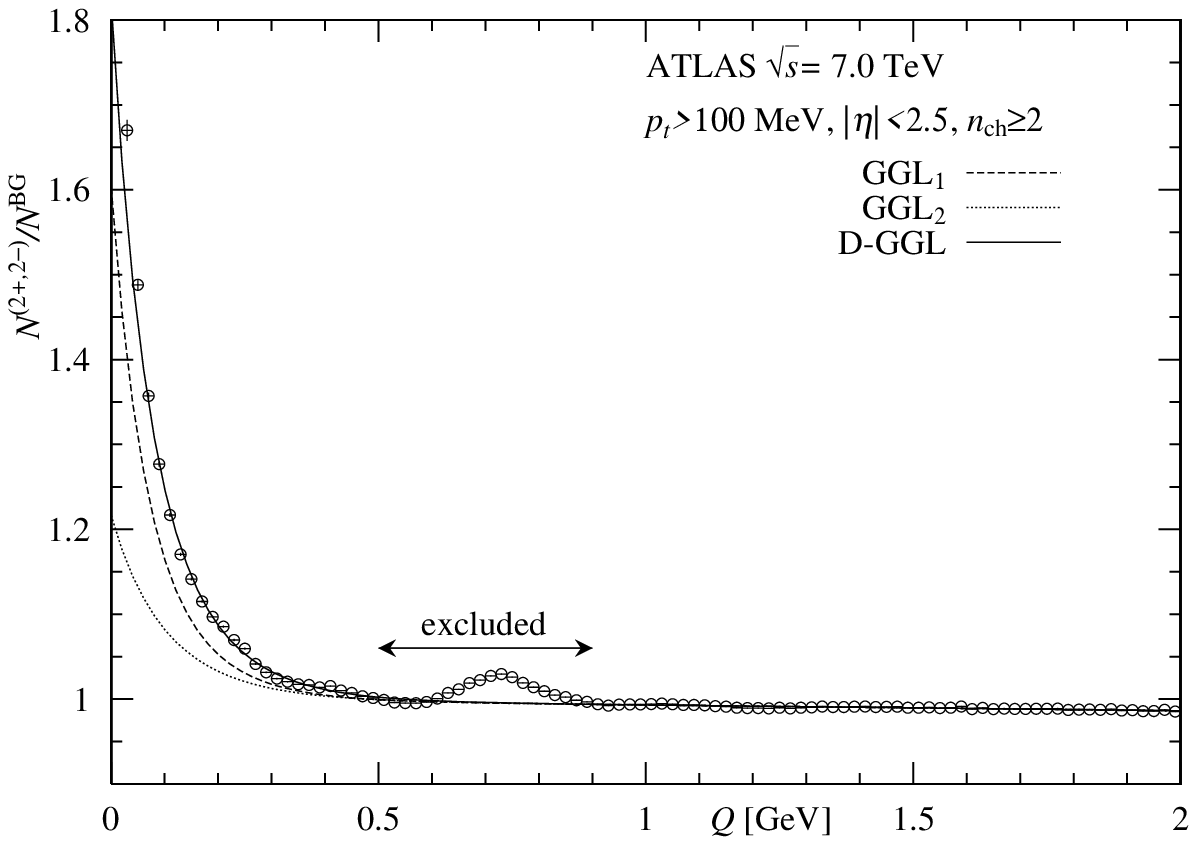}
  \includegraphics[width=0.48\columnwidth]{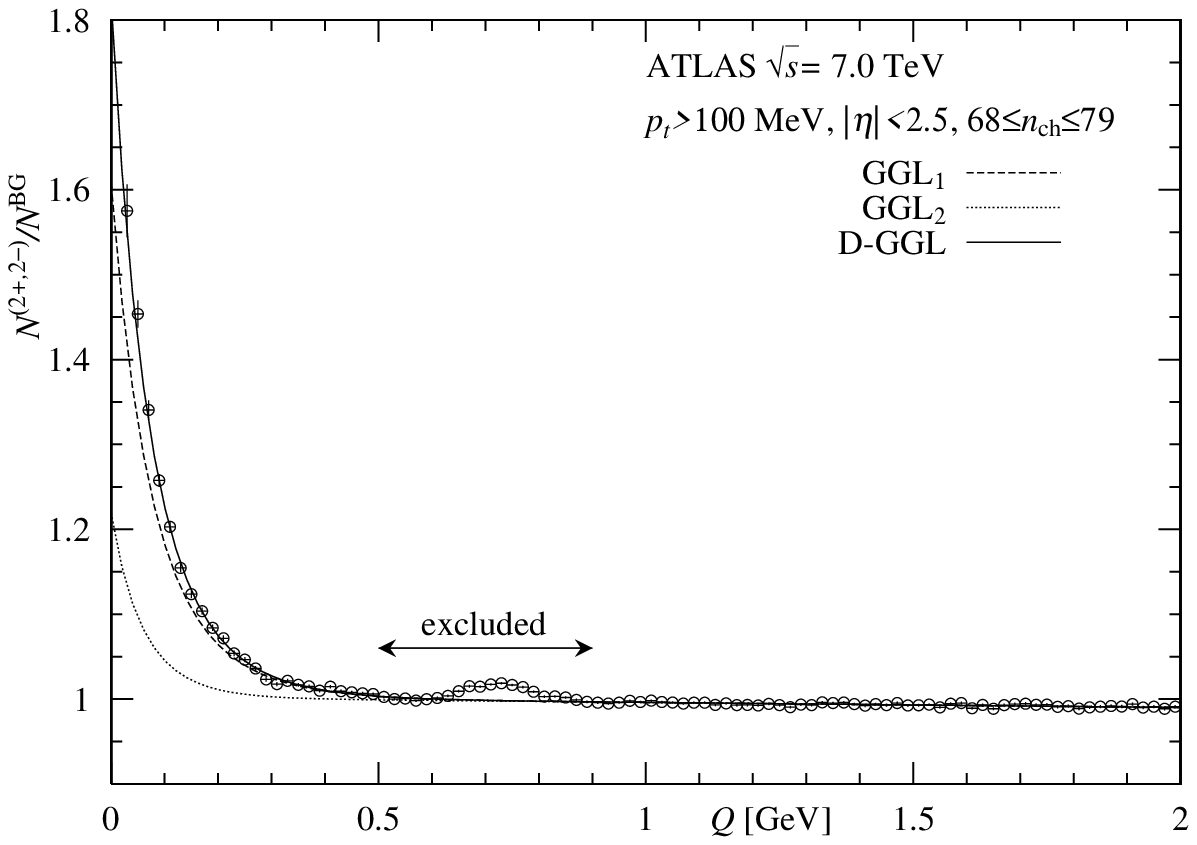}\\
  \includegraphics[width=0.48\columnwidth]{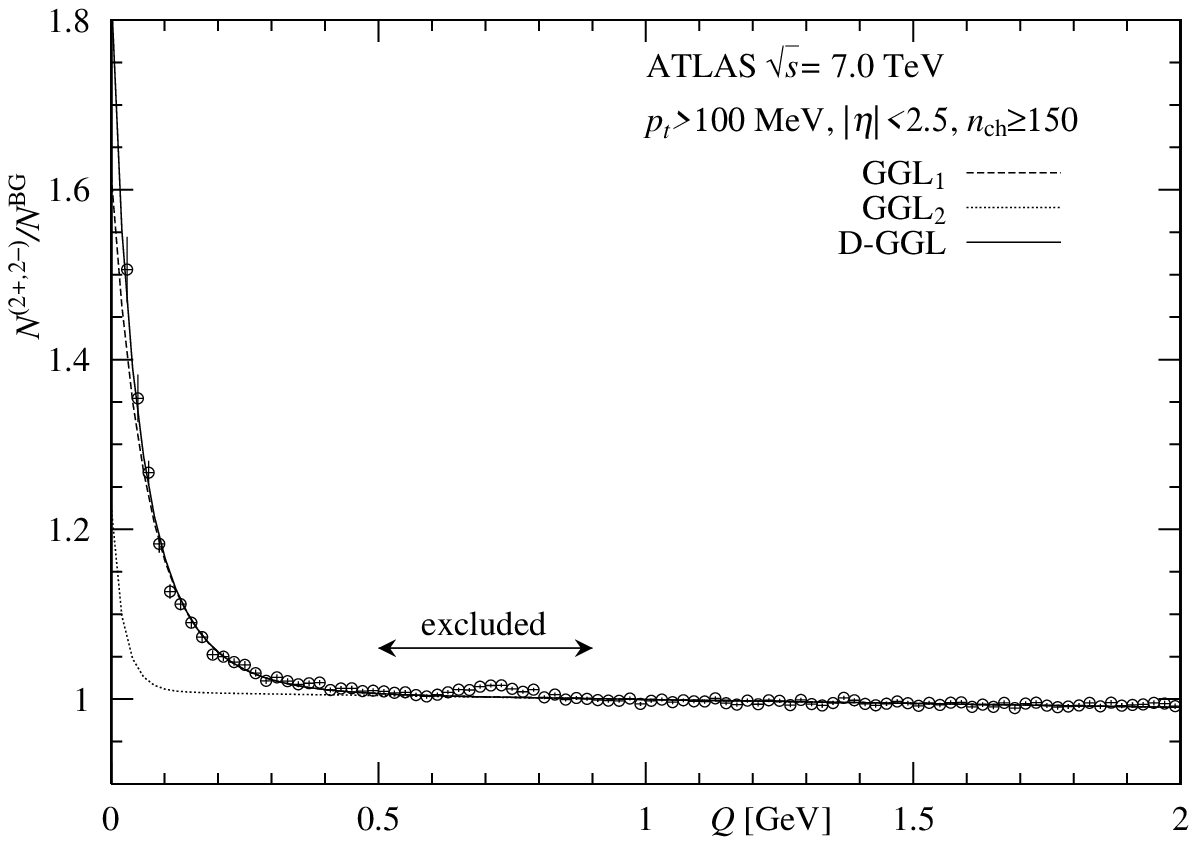}
  \caption{\label{fig5}Analyses of ATLAS data on the BEC ($p_t>100$ MeV, $|\eta| < 2.5$, $n_{\rm ch} \ge 2$) using Eq. (\ref{eq8}) with $E_{\rm BE} = \exp(-RQ)$.}
\end{figure}

By employing Eqs. (\ref{eq8}) and (\ref{eq9}), we can analyze data on the Bose--Einstein correlation at $\sqrt s =$ 0.9 and 7.0 TeV \cite{Aad:2015sja,Astalos:2015zzp}. Our results are presented in Figs. \ref{fig4} and \ref{fig5} and Table \ref{tab2}. Here, $R$ denotes the magnitude of the interaction region in the exponential formula.


\begin{table}[htbp]
\centering
\caption{\label{tab2}Analyses of Bose--Einstein correlation (BEC) collected by the ATLAS collaboration ($p_t>100$ MeV, $|\eta| < 2.5$, $n_{\rm ch} \ge 2$) using Eqs. (\ref{eq9}), (\ref{eq8}), and (\ref{eq10}). Here, $\tilde{\alpha}$ is a new weight factor for data with the interval ($\Delta n$). Physical quantities, denoted with bars ($\bar{\alpha}$, $\bar k_1$, $\bar k_2$, $\bar p_1$, and $\bar p_2$) are taken from Table \ref{tab1}. The formulas of BEC contain the long range correlation as $(1+\varepsilon Q)$.}
\vspace{1mm}
\begin{tabular}{c|cccccccc}
\hline
BEC & \multicolumn{8}{c}{Eq. (\ref{eq9})\quad D-NBD}\\
$\sqrt s$ [TeV] & $n$ & $\bar{\alpha}$, $\tilde{\alpha}$ & $\bar k_1$ & $\bar k_2$ & $R_1$ [fm] & $R_2$ [fm] & $\varepsilon$ ($\times 10^{-3}$) & $\chi^2$/n.d.f. \\
\hline
  \lw{0.9}
& $n\ge 2$
& 0.23
& 7.52
& 2.70
& 1.7$\pm$0.4
& 1.7$\pm$0.1
& 1.5$\pm$1.8
& 98.6/75 \\

& $36\le n\le 45$
& 0.81
& 7.52
& 2.70
& 1.5$\pm$0.2
& 1.5$\pm$0.3
& $-$2.1$\pm$4.1
& 55.8/75 \\
\hline
& $n\ge 2$
& 0.60
& 2.61
& 2.72
& 1.8$\pm$0.01
& 3.1$\pm$0.1
& $-$7.1$\pm$0.2
& 743/75 \\

  7.0
& $68\le n\le 79$
& 1.00
& 2.61
& ---
& 2.4$\pm$0.0
& ---
& $-$8.1$\pm$0.1
& 138/76 \\

& $n\ge 150$
& 1.00
& 2.61
& ---
& 3.0$\pm$0.0
& ---
& $-$12.5$\pm$0.2
& 161/76 \\
\hline
& \multicolumn{8}{c}{Eq. (\ref{eq8})\quad D-GGL}\\
$\sqrt s$ [TeV] & $n$ & $\bar{\alpha}$, $\tilde{\alpha}$ & $\bar p_1$ & $\bar p_2$ & $R_1$ [fm] & $R_2$ [fm] & $\varepsilon$ ($\times 10^{-3}$) & $\chi^2$/n.d.f. \\
\hline
  \lw{0.9}
& $n\ge 2$
& 0.41
& 0.25
& 0.39
& 2.8$\pm$0.2
& 2.7$\pm$0.1
& 3.5$\pm$1.9
& 148/75 \\

& $36\le n\le 45$
& 0.965
& 0.25
& 0.39
& 3.1$\pm$0.2
& 2.9$\pm$2.0
& $-$1.9$\pm$4.1
& 47.2/75 \\
\hline
& $n\ge 2$
& 0.67
& 0.68
& 0.42
& 3.6$\pm$0.1
& 3.2$\pm$0.1
& $-$6.7$\pm$0.3
& 629/75 \\

  7.0
& $68\le n\le 79$
& 1.00
& 0.68
& ---
& 3.8$\pm$0.0
& ---
& $-$6.8$\pm$0.1
& 104/76 \\

& $n\ge 150$
& 1.00
& 0.68
& ---
& 4.9$\pm$0.1
& ---
& $-$11.5$\pm$0.2
& 132/76 \\
\hline
\end{tabular}
\vspace*{1mm}\\
\begin{tabular}{c|ccccc}
\hline
BEC & \multicolumn{5}{c}{Eq. (\ref{eq10})\quad conventional formula}\\
$\sqrt s$ [TeV] & $n$ & $\lambda_{\rm eff}$ & $R$ & $\varepsilon$ ($\times 10^{-3}$) & $\chi^2/$n.d.f.\\
\hline
  \lw{0.9}
& $n\ge 2$
& 0.74$\pm$0.03
& 1.8$\pm$0.1
& $-$0.19$\pm$1.86
& 86.0/75 \\

& $36\le n\le 45$
& 0.75$\pm$0.12
& 2.3$\pm$0.2
& $-$5.8$\pm$3.9
& 33.9/75 \\
\hline
& $n\ge 2$
& 0.72$\pm$0.01
& 2.1$\pm$0.0
& $-$8.3$\pm$0.2
& 919/75 \\

  7.0
& $68\le n\le 79$
& 0.72$\pm$0.02
& 2.3$\pm$0.0
& $-$7.6$\pm$0.6
& 133/75 \\

& $n\ge 150$
& 0.53$\pm$0.03
& 2.4$\pm$0.1
& $-$10.2$\pm$0.9
& 125/75 \\
\hline
\end{tabular}
\end{table}

Combining the results from Tables \ref{tab1} and \ref{tab2}, we can choose favorable frameworks that govern the multiplicity distribution $P(n)$ and Bose--Einstein correlations at $\sqrt s =$0.9 and 7 TeV for the ATLAS collaboration data. For the BEC with the constraints of multiplicity, by taking into account the averaged probabilities over the interval ($\Delta n$) and  calculating new weight factors (denoted as $\tilde{\alpha}$) we can analyze data on the BEC with multiplicity intervals.


\section{\label{sec4}Analyses of MD and the BEC at $\sqrt s =$ 0.9, and 7 TeV for CMS data}

\begin{figure}[htbp]
  \centering
  \includegraphics[width=0.48\columnwidth]{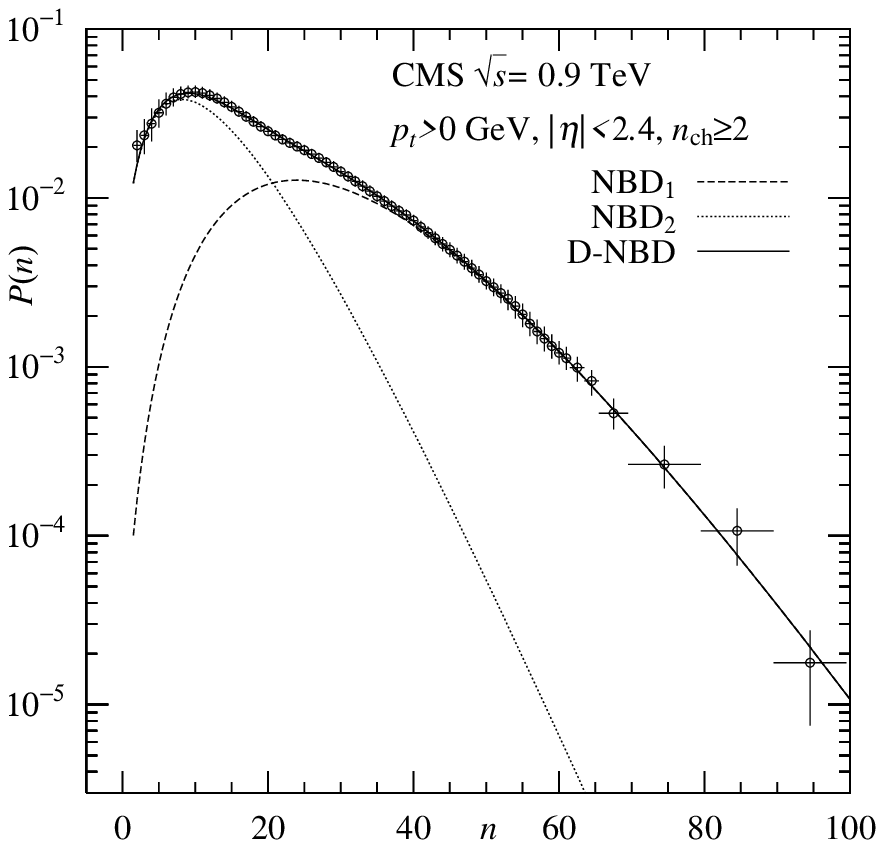}
  \includegraphics[width=0.48\columnwidth]{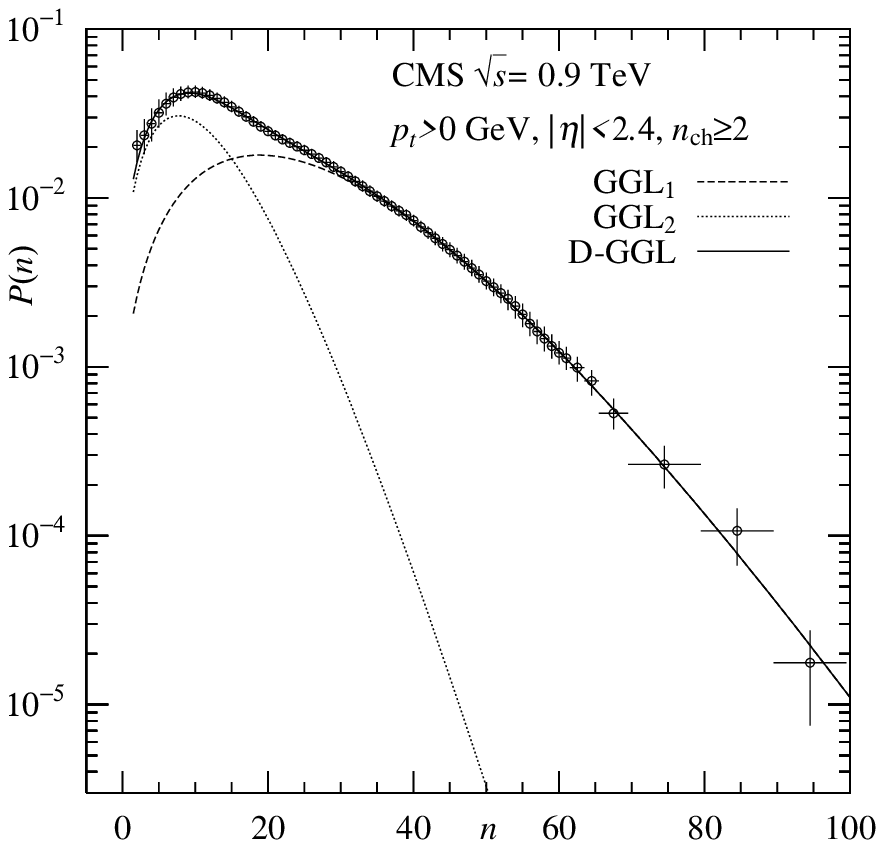}\\
  \includegraphics[width=0.48\columnwidth]{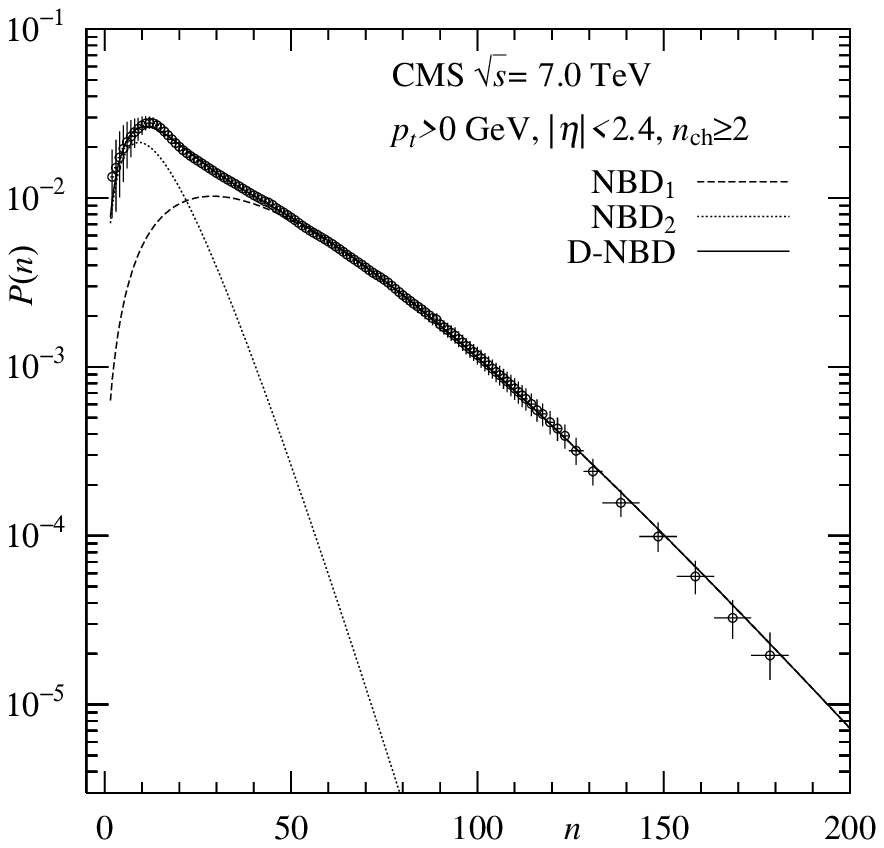}
  \includegraphics[width=0.48\columnwidth]{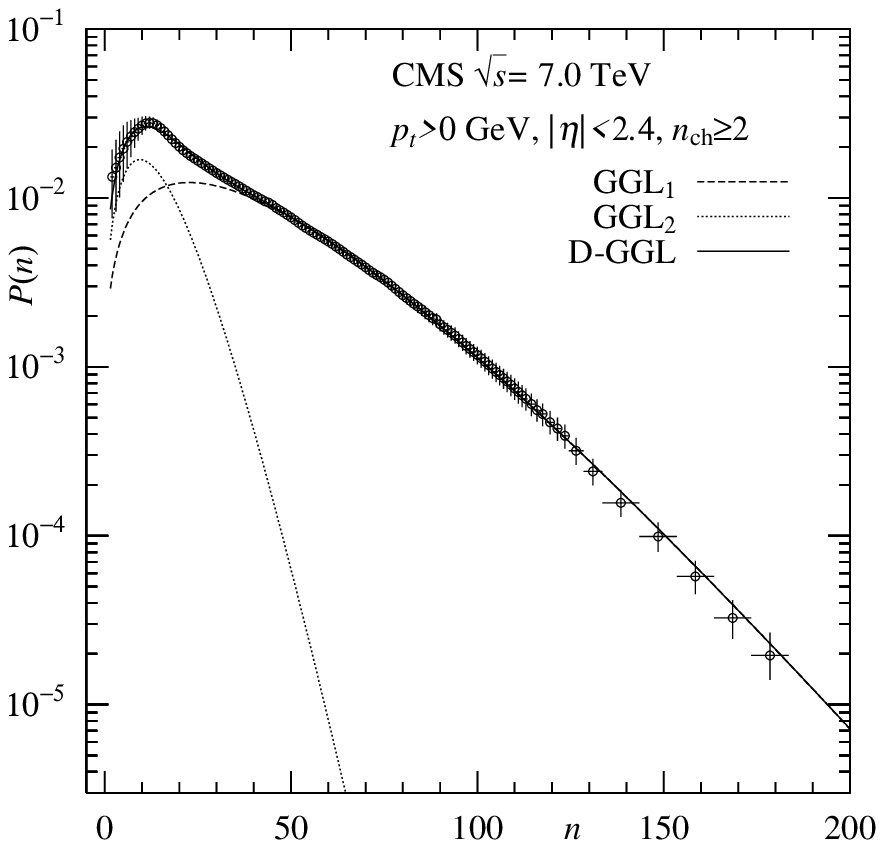}
  \caption{\label{fig6}Analyses of CMS data ($p_t>0$ GeV, $|\eta| < 2.4$, $n_{\rm ch} \ge 2$) using Eqs. (\ref{eq1}) and (\ref{eq5}).}
\end{figure}

\begin{figure}[htbp]
  \centering
  \includegraphics[width=0.48\columnwidth]{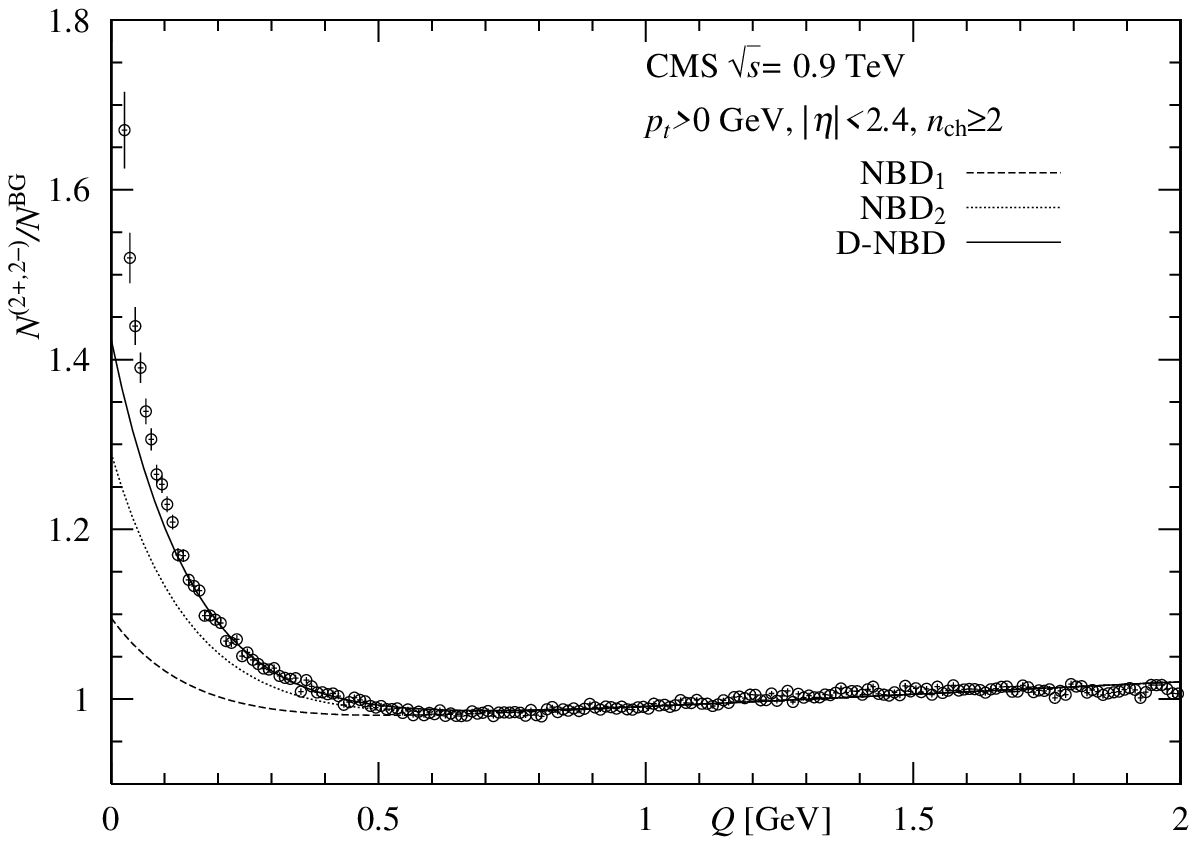}
  \includegraphics[width=0.48\columnwidth]{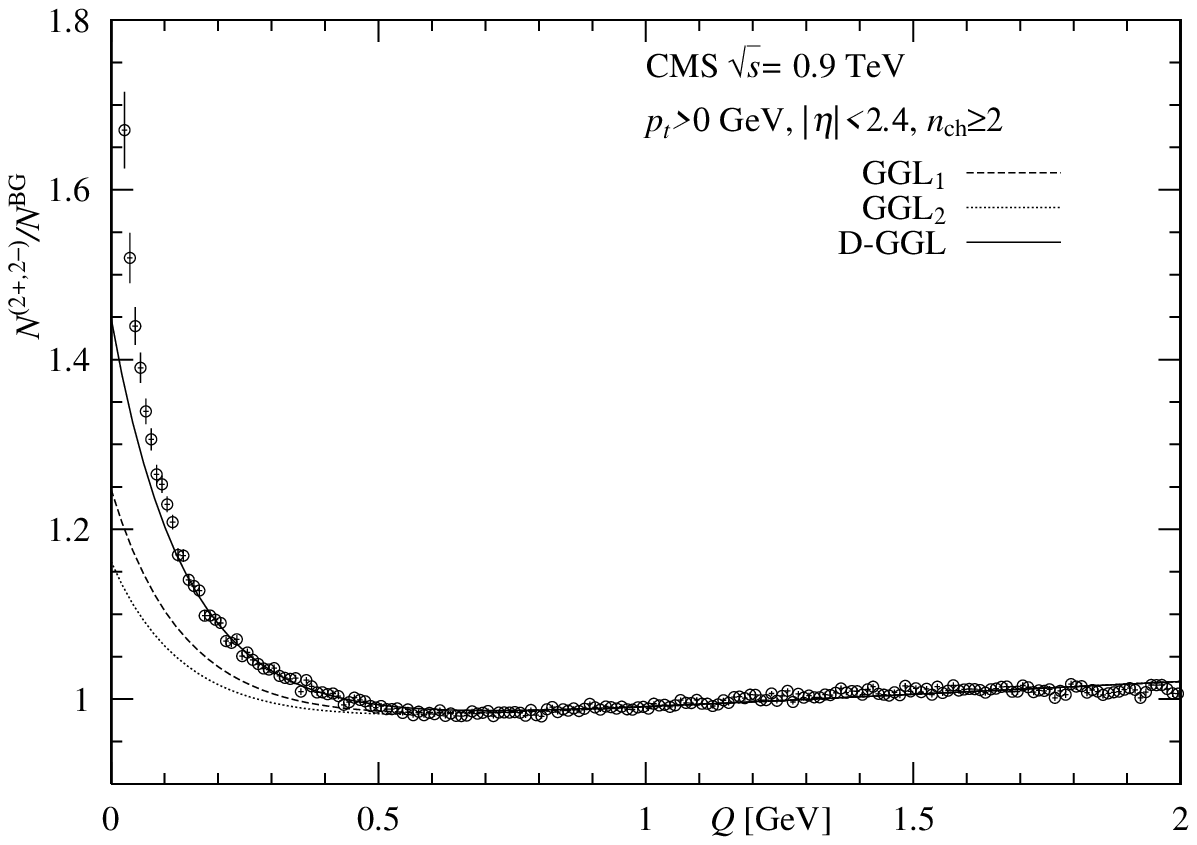}\\
  \includegraphics[width=0.48\columnwidth]{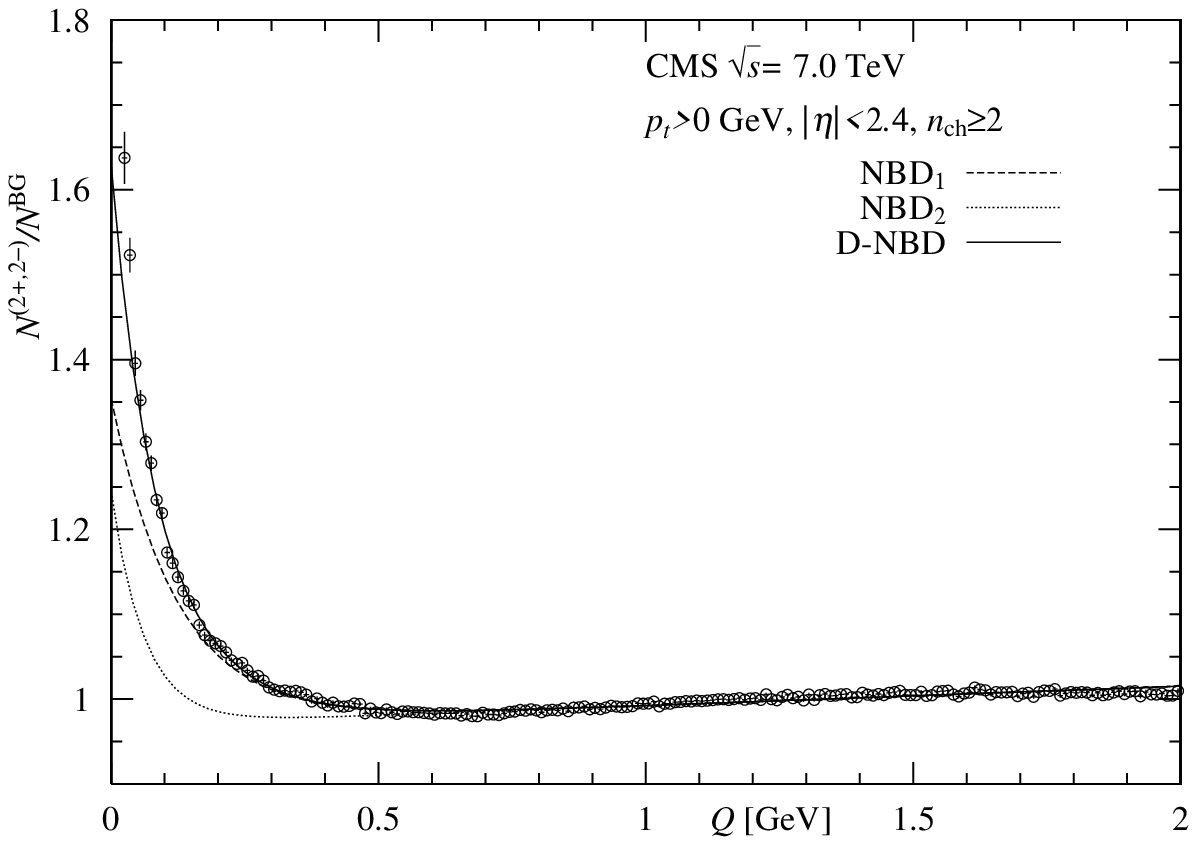}
  \includegraphics[width=0.48\columnwidth]{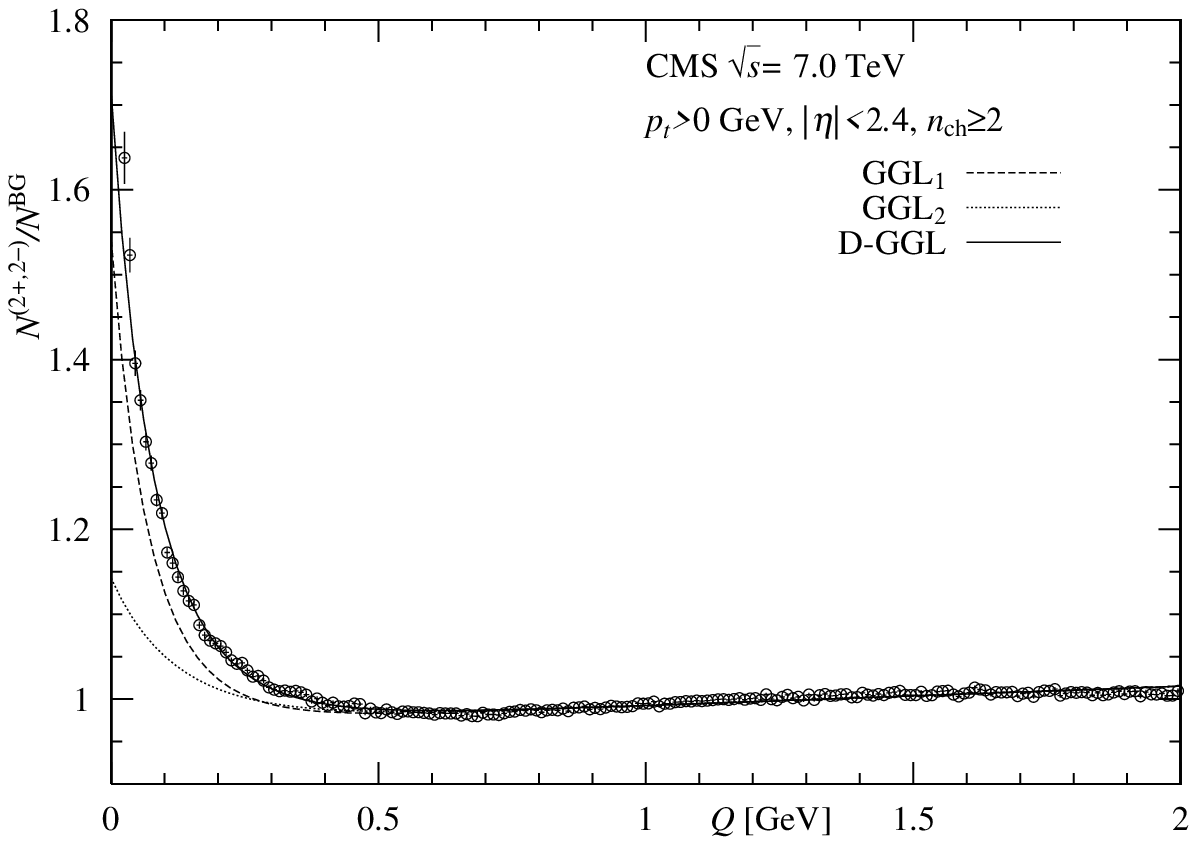}
  \caption{\label{fig7}Analyses of CMS data ($p_t>0$ GeV, $|\eta| < 2.4$, $n_{\rm ch} \ge 2$) using Eqs. (\ref{eq9}) and (\ref{eq8}).}
\end{figure}

In this section, we present our analyses of MD for data collected by the CMS collaboration \cite{Khachatryan:2010nk}, using Eqs. (\ref{eq1}) and (\ref{eq5}), and the BEC \cite{Khachatryan:2011hi} using Eqs. (\ref{eq8}) and (\ref{eq9}). In our analysis of the data on MD ($P(n)$), as in the ATLAS case, $P(0)$ and $P(1)$ are disregarded, because the estimated parameters are used in the analysis of the data on the BEC. We have adopted a renormalization scheme in our calculations. Our results are presented in Figs. \ref{fig6} and \ref{fig7} and Tables \ref{tab3} and \ref{tab4}. 


\begin{table}[htbp]
\centering
\caption{\label{tab3}Analyses of MD ($P(n)$) collected by the CMS collaboration ($p_t>0$ GeV, $|\eta| < 2.4$, $n_{\rm ch} \ge 2$) using Eqs. (\ref{eq1}) and (\ref{eq5}).}
\vspace{1mm}
\begin{tabular}{c|cccccc}
\hline
MD ($P(n)$) & \multicolumn{6}{c}{Eq. (\ref{eq1})\quad D-NBD}\\
$\sqrt s$ [TeV] & $\alpha$ & $k_1$ & $\langle n_1\rangle$ & $k_2$ & $\langle n_2\rangle$ & $\chi^2/$n.d.f.\\
\hline
 0.9
& 0.40$\pm$0.15
& 5.72$\pm$1.68
& 29.6$\pm$3.8
& 3.52$\pm$0.68
& 12.1$\pm$1.8
& 4.55/61\\

 7.0
& 0.58$\pm$0.06
& 2.94$\pm$0.34
& 44.6$\pm$2.5
& 2.94$\pm$0.43
& 14.7$\pm$1.0
& 13.0/120\\
\hline
& \multicolumn{6}{c}{Eq. (\ref{eq5})\quad D-GGL}\\
$\sqrt s$ [TeV] & $\alpha$ & $p_1$ & $\langle n_1\rangle$ & $p_2$ & $\langle n_2\rangle$ & $\chi^2/$n.d.f.\\
\hline
 0.9
& 0.57$\pm$0.12
& 0.31$\pm$0.08
& 25.3$\pm$2.4
& 0.29$\pm$0.07
& 10.6$\pm$0.9
& 3.95/61\\

 7.0
& 0.69$\pm$0.04
& 0.61$\pm$0.07
& 40.2$\pm$1.2
& 0.35$\pm$0.08
& 13.6$\pm$0.7
& 12.9/120\\
\hline
\end{tabular}
\end{table}


\begin{table}[htbp]
\centering
\caption{\label{tab4}Analyses of the Bose--Einstein correlation (BEC) for CMS collaboration data ($p_t>0$ GeV, $|\eta| < 2.4$) using Eqs. (\ref{eq9}), (\ref{eq8}), and (\ref{eq10}). The formulas of BEC contain the long range correlation as $(1+\varepsilon Q)$.} 
\vspace{1mm}
\begin{tabular}{c|ccccccc}
\hline
BEC & \multicolumn{7}{c}{Eq. (\ref{eq9})\quad D-NBD}\\
$\sqrt s$ [TeV] &  $\bar{\alpha}$ & $\bar k_1$ & $\bar k_2$ & $R_1$ [fm] & $R_2$ [fm] & $\varepsilon$ ($\times 10^{-2}$) & $\chi^2$/n.d.f. \\
\hline
 0.9
& 0.40
& 5.72
& 3.52
& 1.3$\pm$0.1
& 1.3$\pm$0.0
& 3.1$\pm$0.1
& 681/194\\

 7.0
& 0.58
& 2.94
& 2.94
& 1.6$\pm$0.0
& 3.2$\pm$0.1
& 2.4$\pm$0.0
& 691/194\\
\hline
& \multicolumn{7}{c}{Eq. (\ref{eq8})\quad D-GGL}\\
$\sqrt s$ [TeV] & $\bar{\alpha}$ & $\bar p_1$ & $\bar p_2$ & $R_1$ [fm] & $R_2$ [fm] & $\varepsilon$ ($\times 10^{-2}$) & $\chi^2$/n.d.f. \\
\hline
 0.9
& 0.57
& 0.31
& 0.29
& 2.5$\pm$0.1
& 2.5$\pm$0.1
& 3.1$\pm$0.1
& 668/194\\

 7.0
& 0.69
& 0.62
& 0.35
& 3.9$\pm$0.1
& 2.7$\pm$0.1
& 2.4$\pm$0.1
& 683/194\\
\hline
\end{tabular}
\vspace*{1mm}\\
\begin{tabular}{c|cccc}
\hline
BEC & \multicolumn{4}{c}{Eq. (\ref{eq10})\quad conventional formula}\\
$\sqrt s$ [TeV] & $\lambda_{\rm eff}$ & $R$ & $\varepsilon$ ($\times 10^{-2}$) & $\chi^2/$n.d.f.\\
\hline
 0.9
& 0.62$\pm$0.01
& 1.56$\pm$0.02
& 2.8$\pm$0.1  
& 487/194\\
 7.0
& 0.62$\pm$0.01
& 1.9$\pm$0.0
& 2.2$\pm$0.0  
& 738/194\\
\hline
\end{tabular}
\end{table}


\section{\label{sec5}Concluding remarks and discussion}

\begin{figure}[htbp]
  \centering
  \includegraphics[width=0.40\columnwidth]{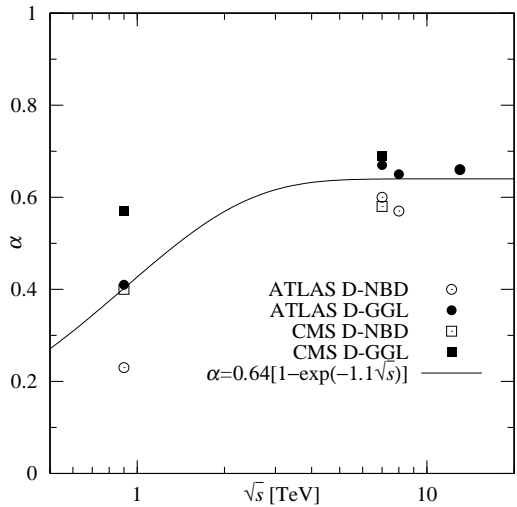}
  \caption{\label{fig8}The energy dependence of the $\alpha$ parameters in Eqs. (\ref{eq1}) and (\ref{eq5}).}
\end{figure}

\paragraph{C1)} As seen in Fig. \ref{fig8}, the $\alpha$ values are almost constant at $\sqrt s \ge 7.0$ -- 8.0 TeV. On the other hand, fluctuations are observed for the $\alpha$ values at $\sqrt s \ge 0.9$ TeV. This behavior represents the beginning of the violation of KNO scaling, because $z_1 = n/\langle n_1\rangle = n/(a_1\langle n\rangle) = z/a_1$ ($a_1 = \langle n_1\rangle/\langle n\rangle$) and $z_2 = n/\langle n_2\rangle = z/a_2$ ($a_2 = \langle n_2\rangle/\langle n\rangle$) in the KNO scaling function in Eq. (\ref{eq1}).
\begin{eqnarray}
  \psi(z) &\!\!\!=&\!\!\! \lim_{n,\, \langle n\rangle \to \infty} \langle n\rangle\left\{ \alpha P_{\rm NBD_1}(n,\,k_{\rm N_1},\,\langle n_1\rangle) + (1-\alpha)P_{\rm NBD_1}(n,\,k_{\rm N_2},\,\langle n_2\rangle)\right\}\nonumber\\
&\!\!\!=&\!\!\! \alpha \frac{k_1^{k_1}}{\Gamma(k_1)}\left(\frac z{a_1}\right)^{k_1-1}e^{-k_1\frac z{a_1}} + (1-\alpha)\frac{k_2^{k_2}}{\Gamma(k_2)}\left(\frac z{a_2}\right)^{k_2-1}e^{-k_2\frac z{a_2}}
\label{eq11}
\end{eqnarray}
Because $\psi(z)$ contains $\alpha$, $k_1$, $k_2$, $a_1$, and $a_2$, the violation of KNO scaling is obvious.


\paragraph{C2)} It can be inferred from the analyses of the data on MD ($P(n)$) and the BEC that the D-GGL formula performs as effectively as the D-NBD. See Tables \ref{tab1}$\sim$\ref{tab4}. This can also be seen from a somewhat different point of view. Rewriting $\kappa =2/k_{\rm N}$ in the NBD as in Fig. \ref{fig1}, we are able to compare it with $p$ in the GGL formula as
\begin{eqnarray}
\kappa = 1/[ 1 + (k_{\rm N} - 2)/2] \Longleftrightarrow p=1/( 1 + \gamma).
\label{eq12}
\end{eqnarray}
Our calculations are depicted in Table \ref{tab5}. The corresponding relation (\ref{eq12}) appears to be satisfied provided that $\langle n_1\rangle \gg 1$. We can say that the NBD and the GGL formula are complementary to each other.


\begin{table}[htbp]
\centering
\caption{\label{tab5}Comparison of $\kappa_i=2/k_{\rm N_i}$ from D-NBD with $p_i$ from the D-GGL formula.} 
\vspace{1mm}
\begin{tabular}{ccccc}
\hline
\multicolumn{5}{c}{MD ($P(n)$) by ATLAS collaboration}\\
$\sqrt s$ [TeV] & $\kappa_1=2/k_{\rm N_1}$ & $p_1$ & $\kappa_2=2/k_{\rm N_2}$ & $p_2$ \\
\hline
0.9
& 0.27$\pm$0.03
& 0.25$\pm$0.04
& 0.74$\pm$0.02
& 0.39$\pm$0.02\\

7.0
& 0.77$\pm$0.03
& 0.68$\pm$0.03
& 0.74$\pm$0.02
& 0.42$\pm$0.02\\

8.0
& 0.74$\pm$0.03
& 0.66$\pm$0.03
& 0.78$\pm$0.03
& 0.45$\pm$0.03\\
13
& 1.00$\pm$0.01
& 0.90$\pm$0.06
& 0.81$\pm$0.02
& 0.60$\pm$0.03\\
\hline
\multicolumn{5}{c}{MD ($P(n)$) by CMS collaboration}\\
$\sqrt s$ [TeV] & $\kappa_1$ & $p_1$ & $\kappa_2$ & $p_2$ \\
\hline
 0.9
& 0.35$\pm$0.10
& 0.31$\pm$0.08
& 0.57$\pm$0.11
& 0.29$\pm$0.07\\

 7.0
& 0.68$\pm$0.08
& 0.61$\pm$0.07
& 0.68$\pm$0.10
& 0.35$\pm$0.08\\
\hline
\end{tabular}
\end{table}

\paragraph{C3)} In our analyses of BEC, we have employed parameters estimated in the analyses of MD ($P(n)$). Our values obtained for the $\chi^2$'s in Tables \ref{tab2} and \ref{tab4} show that the adopted procedure indeed seems to be valid. Our results on BEC are summarized in Table \ref{tab6} . Based on values of $\chi^2$'s at 0.9 TeV, eq. (\ref{eq10}) (CF) seems to be suitable at 0.9 TeV. However, concerning $\chi^2$'s at 7 TeV the situation is reversed. Eq. (\ref{eq8}) based on D-GGL seems to be a fairly good description. Of course, in the future, we will have to elucidate those origins of phenomena at LHC.


\begin{table}[htbp]
\centering
\caption{\label{tab6}Summary of Tables \ref{tab2} and \ref{tab4}.}
\vspace{1mm}
\begin{tabular}{ccc}
\hline
\multicolumn{3}{c}{ATLAS BEC: $\chi^2/$n.d.f.(75)}\\
formula & $\sqrt s = 0.9$ TeV & $\sqrt s = 7.0$ TeV\\
\hline
 CF Eq. (\ref{eq10}) &  86 & 919\\
D-NBD Eq. (\ref{eq9})  &  99 & 743\\
D-GGL Eq. (\ref{eq8})  & 148 & 629\\
\hline
\multicolumn{3}{c}{CMS BEC: $\chi^2/$n.d.f.(194)}\\
formula & $\sqrt s = 0.9$ TeV & $\sqrt s = 7.0$ TeV\\
\hline
 CF Eq. (\ref{eq10}) & 487 & 738\\
D-NBD Eq. (\ref{eq9})  & 681 & 691\\
D-GGL Eq. (\ref{eq8})  & 668 & 683\\
\hline
\end{tabular}
\end{table}

\begin{figure}[htbp]
  \centering
  \includegraphics[width=0.40\columnwidth]{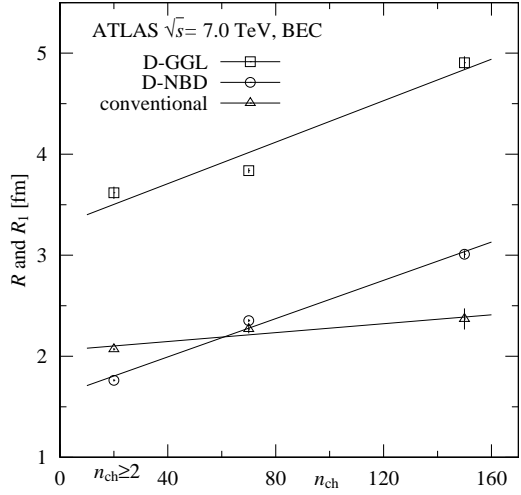}
  \caption{\label{fig9}The multiplicity ($n_{\rm ch}$) dependences of $R$ and $R_1$ in Eqs. (\ref{eq8}), (\ref{eq9}), and (\ref{eq10}) for the BEC.}
\end{figure}

\paragraph{C4)} The ATLAS collaboration has stressed that the interaction ranges estimated in their analyses of the data on the BEC demonstrate the saturation of the interaction ranges ($R$'s) as the multiplicity increases when the conventional formula is utilized. However, our results on the BEC for the ATLAS collaboration data have shown a different description, i.e., the interaction range increases as the multiplicity increases. See, Fig. \ref{fig9}.


\paragraph{D1)} Our two-component model is necessary to introduce two corresponding sources or two kinds of collision mechanism \cite{GrosseOetringhaus:2009kz,Navin:2010kk}. The following correspondences may be inferred, because $\langle n_1\rangle > \langle n_2\rangle$:
\begin{eqnarray}
\left\{
\begin{array}{l}
  P_1 \Longleftrightarrow \mbox{Non-diffractive (ND) process},\medskip\\
  P_2 \Longleftrightarrow 
\left\{
\begin{array}{l}
\mbox{single diffractive (SD) process},\medskip\\
\mbox{double diffractive (DD) process}.\medskip\\
\end{array}
\right.
\end{array}
\right.
\label{eq13}
\end{eqnarray}
The weight factor $\alpha$ in Eqs. (\ref{eq1}) and (\ref{eq5}) can be interpreted by means of various cross sections, $\sigma_{\rm ND}$, $\sigma_{\rm 2SD}$, $\sigma_{\rm DD}$, and $\sigma_{\rm inel} = \sigma_{\rm ND}+\sigma_{\rm 2SD}+\sigma_{\rm DD}$, as
\begin{eqnarray}
\left\{
\begin{array}{l}
  \ \quad \alpha \quad \Longleftrightarrow \sigma_{\rm ND}/\sigma_{\rm inel},\medskip\\
  (1-\alpha) \Longleftrightarrow (\sigma_{\rm 2SD}+\sigma_{\rm DD})/\sigma_{\rm inel}.\\
\end{array}
\right.
\label{eq14}
\end{eqnarray}
In other words, $\alpha$ means the occurrence rate in two kinds of collisions \cite{ATLAS:2010mza,Ciesielski:2012mc}.

\begin{figure}[htbp]
  \centering
  \includegraphics[width=0.40\columnwidth]{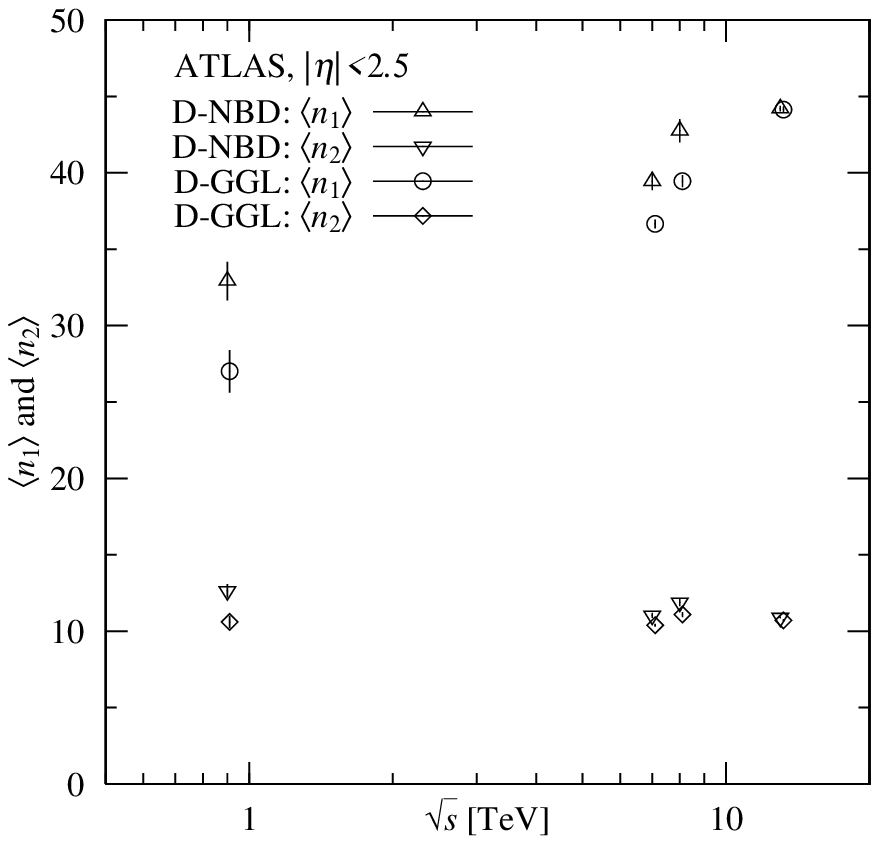}\qquad
  \includegraphics[width=0.40\columnwidth]{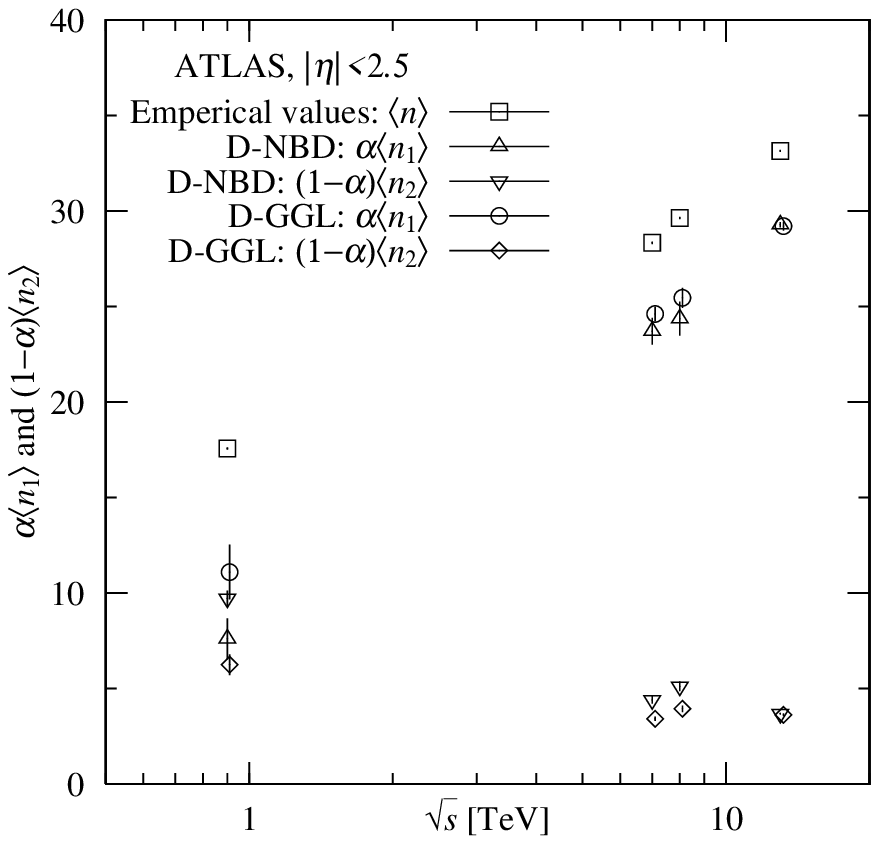}\\
  (a)\hspace{70mm} (b)
  \caption{\label{fig10}(a) The energy dependences of the parameters $\langle n_1\rangle$ and $\langle n_2\rangle$, (b) $\alpha\langle n_1\rangle$ and $(1-\alpha)\langle n_2\rangle$ in Eqs. (\ref{eq1}) and (\ref{eq5}).}
\end{figure}

The total average multiplicity in Eqs. (\ref{eq1}) and (\ref{eq5}) is defined as 
\begin{eqnarray}
  \langle n\rangle = \alpha \sum_n P_1(n,\,\langle n_1\rangle)n + (1-\alpha) \sum_n P_2(n,\,\langle n_2\rangle)n = \alpha \langle n_1\rangle + (1-\alpha) \langle n_2\rangle
\label{eq15}
\end{eqnarray}
Various kinds of average multiplicities $\langle n_1\rangle$, $\langle n_2\rangle$, $\alpha\langle n_1\rangle$, $(1-\alpha)\langle n_2\rangle$ are shown in Fig. \ref{fig10}(a) and (b).


\begin{figure}[htbp]
  \centering
  \includegraphics[width=0.48\columnwidth]{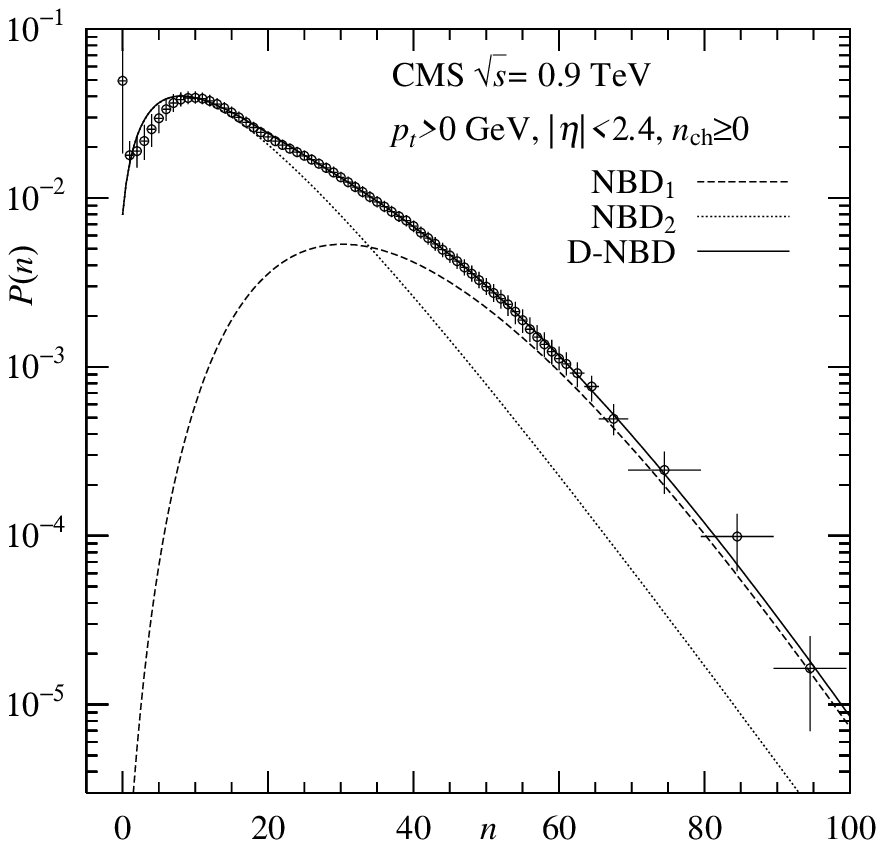}
  \includegraphics[width=0.48\columnwidth]{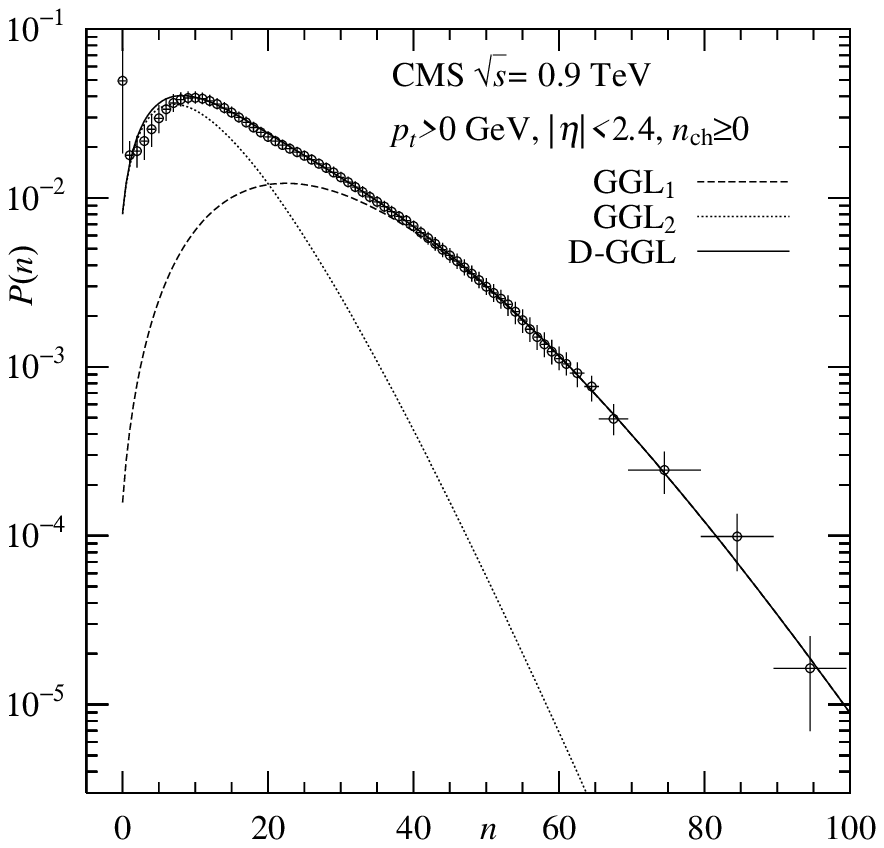}\\
  \includegraphics[width=0.48\columnwidth]{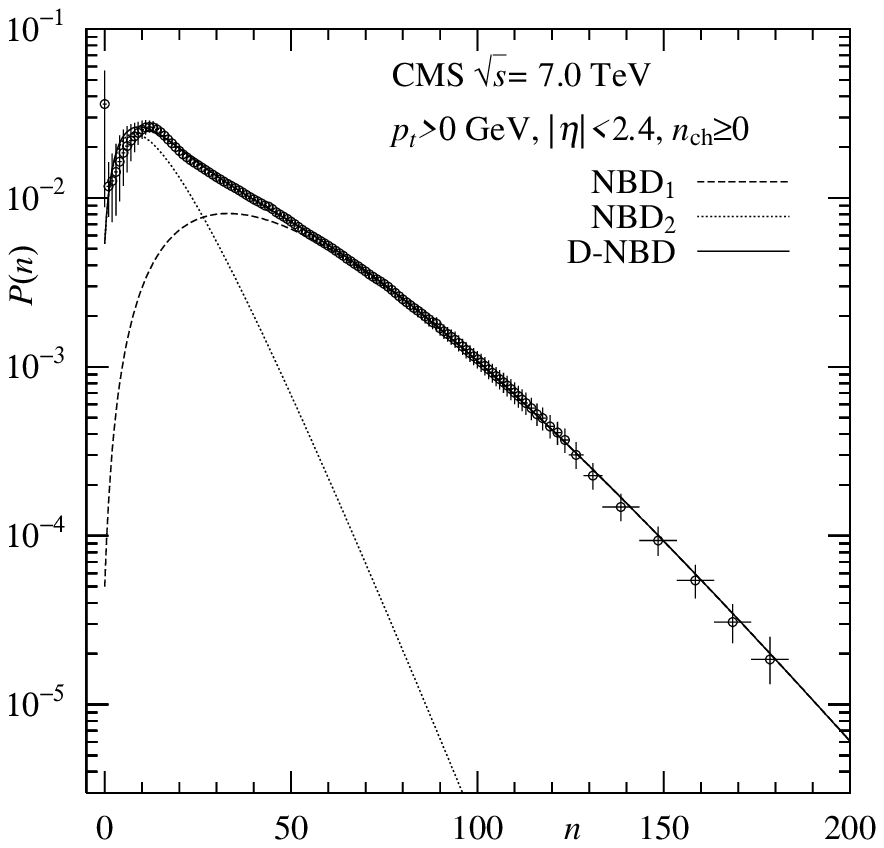}
  \includegraphics[width=0.48\columnwidth]{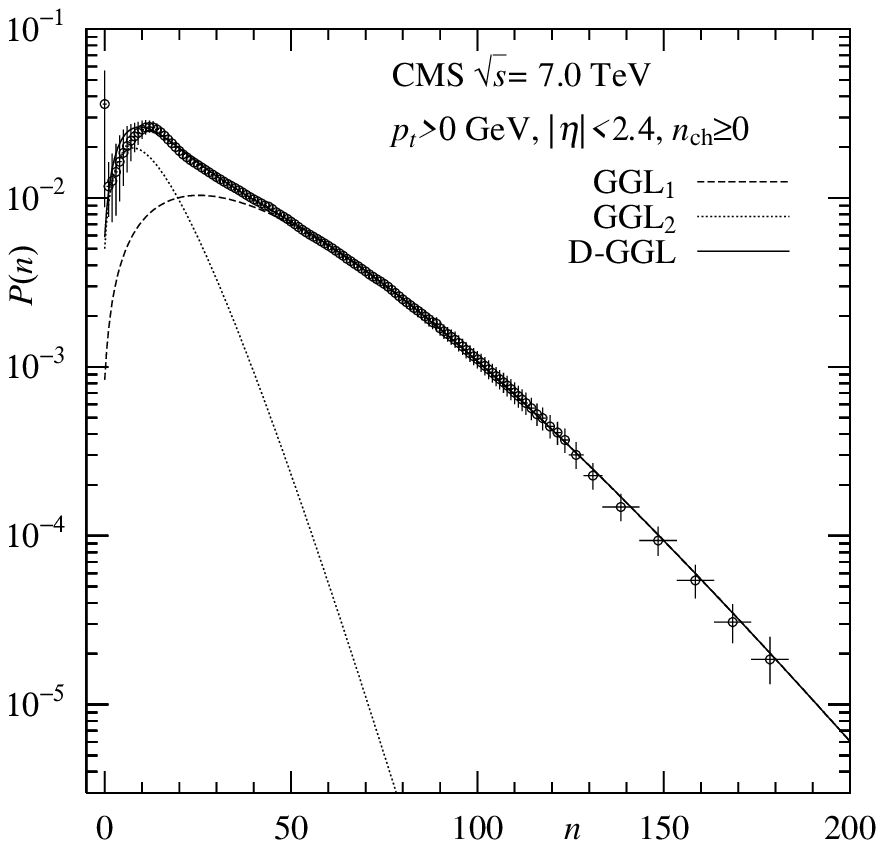}
  \caption{\label{fig11}Analyses of CMS data ($p_t>0$ GeV, $|\eta| < 2.4$, $n_{\rm ch} \ge 0$) using Eqs. (\ref{eq1}) and (\ref{eq5}).}
\end{figure}

\paragraph{D2)} In the fourth section, to investigate BEC by CMS collaboration, we have analyzed data excluding $P(0)$ and $P(1)$ as ATLAS collaboration did. We show our results of analysis of data including $P(0)$ and $P(1)$ using Eqs. (\ref{eq1}) and (\ref{eq5}) in Fig. \ref{fig11} and Table \ref{tab7}. As compared values of $\chi^2$s at 0.9 TeV in Table \ref{tab7} and that in Table \ref{tab3}, the formers are larger than the latters.  However, those at 7.0 TeV are almost the same. In Ref. \cite{Gieseke:2017vli}, the value of $\chi^2/{\rm n.d.f.} = 0.84$ (color reconnection approach in the Monte Carlo study) for the analysis of MD ($P(n)$) at 7.0 TeV is mentioned. Ours in Table \ref{tab7} are compatible with that above.


\begin{table}[htbp]
\centering
\caption{\label{tab7}Analyses of MD ($P(n)$) collected by the CMS collaboration ($p_t>0$ GeV, $|\eta| < 2.4$, $n_{\rm ch} \ge 0$) using Eqs. (\ref{eq1}) and (\ref{eq5}).}
\vspace{1mm}
\begin{tabular}{c|cccccc}
\hline
MD ($P(n)$) & \multicolumn{6}{c}{Eq. (\ref{eq1})\quad D-NBD}\\
$\sqrt s$ [TeV] & $\alpha$ & $k_1$ & $\langle n_1\rangle$ & $k_2$ & $\langle n_2\rangle$ & $\chi^2/$n.d.f.\\
\hline
  0.9
& 0.17$\pm$0.12
& 8.08$\pm$2.60
& 35.1$\pm$4.0
& 2.35$\pm$0.30
& 14.6$\pm$2.1
& 13.9/63\\

  7.0
& 0.47$\pm$0.08
& 3.36$\pm$0.51
& 47.8$\pm$3.6
& 2.22$\pm$0.23
& 15.4$\pm$1.6
& 13.7/122\\
\hline
& \multicolumn{6}{c}{Eq. (\ref{eq5})\quad D-GGL}\\
$\sqrt s$ [TeV] & $\alpha$ & $p_1$ & $\langle n_1\rangle$ & $p_2$ & $\langle n_2\rangle$ & $\chi^2/$n.d.f.\\
\hline
  0.9
& 0.40$\pm$0.16
& 0.24$\pm$0.09
& 28.1$\pm$4.1
& 0.45$\pm$0.08
& 11.5$\pm$1.8
& 12.9/63\\

  7.0
& 0.60$\pm$0.06
& 0.54$\pm$0.08
& 41.9$\pm$2.2
& 0.58$\pm$0.13
& 13.7$\pm$1.2
& 13.4/122\\
\hline
\end{tabular}
\end{table}

\paragraph{D3)} Concerning the triple-NBD formula (T-NBD) proposed in \cite{Zborovsky:2013tla}, it is worthwhile to mention the following:\smallskip\\
1) Comparing $\chi^2$'s by T-NBD with those by D-NBD, the formers are very much smaller than the latter. In other words, the T-NBD seems to be an excellent description.\smallskip\\
2) By making use of estimated values [$\alpha_i$ and $k_i$ ($i=1,\,2,\,3$)] in \cite{Zborovsky:2013tla}, the effective degrees of coherence are computed as, 
\begin{eqnarray*}
  \lambda_{\rm eff} &\!\!\!=&\!\!\! \alpha_1\frac 2{k_1} + \alpha_2\frac 2{k_2} + \alpha_3\frac 2{k_3} \\
&\!\!\!=&\!\!\! 0.824\frac 2{1.66} + 0.107\frac 2{6.6} + O(0) = 1.03\ {\rm at\ 7.0\ TeV\ (CMS\ Coll.)},\\
  \lambda_{\rm eff} &\!\!\!=&\!\!\! 0.737\frac 2{1.50} + 0.183\frac 2{5.7} + O(0) = 1.05\ {\rm at\ 7.0\ TeV\ (ATLAS\ Coll.)}.
\end{eqnarray*}
They are larger than $\lambda_{\rm eff}$'s by Eq.~(\ref{eq10}) [0.62$\pm$0.01 (CMS Coll.) and 0.72$\pm$0.01 (ATLAS Coll.) in Tables \ref{tab2} and \ref{tab4}]. Those calculations depend on our formula Eq.~(\ref{eq9}) which contains three $k$'s, of course. It should be noticed that the large $\lambda_{\rm eff}$'s above are estimated, provided that our formula, Eq. (\ref{eq9}), is applied to the T-NBD.\\

\noindent
{\it Acknowledgements.} MB would like to thank the colleagues of the Department of Physics, Shinshu University for their kindness. Authors wish to express their appreciation for a referee's suggestions. Moreover, they also thank Dr. Edward Sarkisyan-Grinbaum for his kind information.


\appendix

\section{\label{secA}Stochastic background of the NBD and the GGL formula}
Eqs. (\ref{eq2}) and (\ref{eq3}) are solutions of the following stochastic differential-difference (DD) equation \cite{Biyajima:1984ay},
\begin{eqnarray}
\diff{P(n,\,t)}{t} &\!\!\!=&\!\!\! - \lambda_0 [P(n,\,t)-P(n-1,\,t)]
+ \lambda_1 [(n+1)P(n+1,\,t)-nP(n,\,t)]
\nonumber\\
&\!\!\! &\!\!\! + \lambda_2 [(n-1)P(n-1,\,t)-nP(n,\,t)],
\label{eq16}
\end{eqnarray}
where $\lambda_0$, $\lambda_1$, and $\lambda_2$ are parameters and $P(n,\,t)$ is the probability distribution. In Table \ref{tab8}, physical meanings of parameters in Eq. (\ref{eq16}) are shown: $\lambda_1$ is a death rate, $\lambda_2$ is a birth rate, and $\lambda_0$ is an immigration rate per time. 


\begin{table}[htbp]
\centering
\caption{\label{tab8}Physical meanings of $\lambda_0$, $\lambda_1$, and $\lambda_2$.}
\vspace{1mm}
\begin{tabular}{c|ccc}
\hline
variable & $\lambda_0$ & $\lambda_1$ & $\lambda_2$\\
\hline
\vspace{-3mm}\\
$t$ (time) & $s\langle n\rangle$ & $\dfrac{s\langle n\rangle}k + s$ & $\dfrac{s\langle n\rangle}k$\\
\hline
\multicolumn{4}{c}{$\langle n\rangle = \lambda_0/(\lambda_1 - \lambda_2)$, $s = \lambda_1 - \lambda_2$,}\\
\multicolumn{4}{c}{$k=\lambda_0/\lambda_2$, and $p=1-e^{-st}$}\\
\hline
\end{tabular}
\end{table}

In order to solve Eq. (\ref{eq16}), we use the generating function,
\begin{eqnarray}
  Q(w,\,t) = \sum_{n=0}^{\infty} P(n,\,t) w^n
\label{eq17}
\end{eqnarray}
In terms of Eqs. (\ref{eq16}) and (\ref{eq17}), we obtain the following partial differential equation,
\begin{eqnarray}
\diff{Q(w,\,t)}{t} = \lambda_0 (w-1)Q(w,\,t) + (\lambda_2 w-\lambda_1)(w-1) \diff{Q(w,\,t)}{w}
\label{eq18}
\end{eqnarray}
By making use of two kinds of initial conditions in Table \ref{tab9}, we obtain Eqs. (\ref{eq2}) and (\ref{eq3}), provided that $\beta \bar n/\langle n\rangle = 1$. 


\begin{table}[htbp]
\centering
\caption{\label{tab9}Stochastic background of Eqs. (\ref{eq2}) and(\ref{eq3}).}
\vspace{1mm}
\begin{tabular}{c|c}
\hline
initial conditions & $P(n)$\\
\hline
\vspace{-3mm}\\
$\delta_{n,\,0}$ & $\dfrac{\Gamma (n+k)}{\Gamma (n+1)\Gamma (k)}\dfrac{(p\langle n\rangle/k)^n}{(1+p\langle n\rangle/k)^{n+k}}$\\
\vspace{-3mm}\\
\hline
\vspace{-3mm}\\
$\dfrac{(\beta \bar n)^n}{n!}e^{-\beta \bar n}$ & $\dfrac{(p\langle n\rangle/k)^n}{(1+p\langle n\rangle/k)^{n+k}}
\exp\left[-\dfrac{\beta (1-p)\bar n}{1+p\langle n\rangle/k}\right]$\\
\vspace{-3mm}\\
\ri{(coherent state)} & $\times L_n^{(k-1)}\left(-\dfrac{\beta (1-p)\bar n}{(p\langle n\rangle/k)(1+p\langle n\rangle/k)}\right)$\\
\hline
\end{tabular}
\end{table}

The KNO scaling function of the GGL formula, the probability density, is obtained from the Fokker--Planck equation of Eq. (\ref{eq16}) \cite{Biyajima:1983qu,Fellera:1951aa}.


\section{\label{secB}Derivation of the BEC in the two-component model}

Taking into account that $P(n,\,\langle n\rangle)$ in Eq.~(\ref{eq1}) is the probability distribution for the charged particles ensembles ($a=(+)$ and $b=(-)$ indicate the positive and negative charges respectively), we can decompose it into two probability distributions with the labels $a$ and $b$ as,
\begin{eqnarray}
P^{(ab)}(n,\, \langle n\rangle,\, k) = \sum_{n=n_a+n_b} P^{(a)}(n_a,\, \langle n^a\rangle,\, k^{(a)})\times P^{(b)}(n_b,\, \langle n^b\rangle,\, k^{(b)})
\label{eq19}
\end{eqnarray}
where $\langle n^a\rangle = \langle n^b\rangle = \langle n\rangle/2$ and $k^{(a)}=k^{(b)}=k/2$. $P_1(n_1,\, \langle n\rangle_1,\, k_1)$ and $P_2(n_2,\, \langle n\rangle_2,\, k_2)$ in Eq. (\ref{eq1}) can also be decomposed into the same charged particle probability distributions. Combining those, we obtain the following relations: 
\begin{eqnarray}
P^{(a)}(n,\, \langle n^a\rangle,\, k^{(a)}) &\!\!\!=&\!\!\! \alpha P_1^{(a)}(n,\, \langle n^a\rangle_1,\, k_1^{(a)}/2) + (1-\alpha)P_2^{(a)}(n,\, \langle n^a\rangle_2,\, k_2^{(a)}/2),
\label{eq20}\\
P^{(b)}(n,\, \langle n^b\rangle,\, k^{(b)}) &\!\!\!=&\!\!\! \alpha P_1^{(b)}(n,\, \langle n^b\rangle_1,\, k_1^{(b)}/2) + (1-\alpha)P_2^{(b)}(n,\, \langle n^b\rangle_2,\, k_2^{(b)}/2).
\label{eq21}
\end{eqnarray}
By employing Eqs. (\ref{eq19})$\sim$(\ref{eq21}) and calculating the number of pairs in the same charged particle ensembles ($(2a)$ and $(2b)$) and the number of pairs in opposite charged particle ensembles ($(ab)$), we obtain the following relations:
\begin{eqnarray}
N_1^{(2a)} + N_1^{(2b)} &\!\!\!=&\!\!\! \sum_n P_1^{(a)}(n,\, \langle n^a\rangle_1,\, k_1^{(a)}/2)\frac{\langle n(n-1)\rangle_1}2 + \sum_n P_1^{(b)}(n,\, \langle n^b\rangle_1,\, k_1^{(b)}/2)\frac{\langle n(n-1)\rangle_1}2\nonumber\\
&\!\!\!=&\!\!\! \frac 12(\langle n^a(n^a-1)\rangle_1+\langle n^b(n^b-1)\rangle_1),
\label{eq22}\\
N_1^{{\rm BG}(ab)} &\!\!\!=&\!\!\! \left(\sum_n P_1^{(a)}(n,\, \langle n^a\rangle_1,\, k_1^{(a)}/2)n\right)\left(\sum_n P_1^{(b)}(n,\, \langle n^b\rangle_1,\, k_1^{(b)}/2)n\right)\nonumber\\
 &\!\!\!=&\!\!\! \langle n^a\rangle_1\langle n^b\rangle_1.
\label{eq23}
\end{eqnarray}
By taking the following ratio, we obtain the BEC as
\begin{eqnarray}
N_1^{(2a:\,2b)} / N_1^{{\rm BG}(ab)} = \langle n(n-1)\rangle_1 / \langle n\rangle_1^2 = 1+2/k_1\cdot E_{\rm BE_1}.
\label{eq24}
\end{eqnarray}
At the final step, we have introduced the exchange function for the BEC, $E_{\rm BE_1}\ \mapright{\ Q\to 0\ }\ 1$ and $\mapright{\ Q\to \infty\ }\ 0$. In the same manner, we obtain the second formula for the second component
\begin{eqnarray}
N_2^{(2a:\,2b)} / N_2^{{\rm BG}(ab)} = 1+2/k_2\cdot E_{\rm BE_2}.
\label{eq25}
\end{eqnarray}
Combining Eqs. (\ref{eq24}) and (\ref{eq25}) with weight factors $\alpha$ and $(1-\alpha)$, we obtain Eq. (\ref{eq9}) from the introduction,
\begin{eqnarray*}
 N^{(2+:\,2-)}_{\rm D-NBD}/N^{\rm BG} = \alpha (1+2/k_{\rm N_1}\cdot E_{\rm BE_1}) + (1-\alpha)(1+2/k_{\rm N_2}\cdot E_{\rm BE_2}).
\end{eqnarray*}
Using the same method for the GGL formula with $k_{\rm G} = 2.0$ and $p$, Eq. (\ref{eq3}), we can obtain Eq. (\ref{eq8}) with $\alpha$ and $p_i$ ($i=1,\,2$) for the two-component model.



\begin{thebibliography}{99}
\bibitem{Aad:2010ac}
  G.~Aad {\it et al.} [ATLAS Collaboration],
  New J.\ Phys.\  {\bf 13} (2011) 053033.

\bibitem{Aaboud:2016itf}
  M.~Aaboud {\it et al.} [ATLAS Collaboration],
  Eur.\ Phys.\ J.\ C {\bf 76} (2016) no.9, 502.

\bibitem{Aamodt:2010pp}
  K.~Aamodt {\it et al.} [ALICE Collaboration],
  Eur.\ Phys.\ J.\ C {\bf 68} (2010) 345.

\bibitem{Khachatryan:2010nk}
  V.~Khachatryan {\it et al.} [CMS Collaboration],
  JHEP {\bf 1101} (2011) 079.

\bibitem{Ghosh:2012xh}
  P.~Ghosh,
  Phys.\ Rev.\ D {\bf 85} (2012) 054017.

\bibitem{Zaccolo:2015udc} 
  V.~Zaccolo [ALICE Collaboration],
  Nucl.\ Phys.\ A {\bf 956} (2016) 529.


\bibitem{Fuglesang:1989st} 
  C.~Fuglesang,
  La Thuile Multiparticle Dynamics 1989 (1989) 193-210 (World Scientific, Singapore, 1990).

\bibitem{Koba:1972ng} 
  Z.~Koba, H.~B.~Nielsen and P.~Olesen,
  Nucl.\ Phys.\ B {\bf 40} (1972) 317.

\bibitem{Ansorge:1988kn} 
  R.~E.~Ansorge {\it et al.} [UA5 Collaboration],
  Z.\ Phys.\ C {\bf 43}, 357 (1989).

\bibitem{Giovannini:1998zb} 
  A.~Giovannini and R.~Ugoccioni,
  Phys.\ Rev.\ D {\bf 59} (1999) 094020.
%
\bibitem{Giovannini:2003ft} 
For three-component model, see,
  A.~Giovannini and R.~Ugoccioni,
  Phys.\ Rev.\ D {\bf 68} (2003) 034009.

\bibitem{Dremin:2004ts}
  I.~M.~Dremin and V.~A.~Nechitailo,
  Phys.\ Rev.\ D {\bf 70} (2004) 034005.

\bibitem{Zborovsky:2013tla}
  I.~Zborovsky,
  J.\ Phys.\ G {\bf 40} (2013) 055005.

\bibitem{Mizoguchi:2010vc}
  T.~Mizoguchi and M.~Biyajima,
  Eur.\ Phys.\ J.\ C {\bf 70} (2010) 1061.

\bibitem{Biyajima:1982un}
  M.~Biyajima,
  Prog.\ Theor.\ Phys.\  {\bf 69} (1983) 966.
   Addendum: [Prog.\ Theor.\ Phys.\  {\bf 70} (1983) 1468].

\bibitem{Biyajima:1984aq} 
  M.~Biyajima and N.~Suzuki,
  Phys.\ Lett.\  {\bf 143B} (1984) 463.

\bibitem{Biyajima:1990ku}
  M.~Biyajima, A.~Bartl, T.~Mizoguchi, O.~Terazawa and N.~Suzuki,
  Prog.\ Theor.\ Phys.\  {\bf 84} (1990) 931;
   Addendum: [Prog.\ Theor.\ Phys.\  {\bf 88} (1992) 157].

\bibitem{Aad:2015sja}
  G.~Aad {\it et al.} [ATLAS Collaboration],
  Eur.\ Phys.\ J.\ C {\bf 75} (2015) no.10, 466.

\bibitem{Astalos:2015zzp}
  R.~Astalo\u s,
  Dr. Thesis 
  ``Bose-Einstein correlations in 7 TeV proton-proton collisions in the ATLAS experiment,'' 
  (Uiversity of Nijmegen, 2015). 
  It should be noticed that his quantum optical formula has a different exponential form, $E_{\rm BE} = \exp(-2RQ)$.

\bibitem{Khachatryan:2011hi}
  V.~Khachatryan {\it et al.} [CMS Collaboration],
  JHEP {\bf 1105} (2011) 029.

\bibitem{GrosseOetringhaus:2009kz}
  J.~F.~Grosse-Oetringhaus and K.~Reygers,
  J.\ Phys.\ G {\bf 37} (2010) 083001.

\bibitem{Navin:2010kk}
  S.~Navin,
  ``Diffraction in Pythia,''
  LUTP-09-23
  [arXiv:1005.3894 [hep-ph]].

\bibitem{ATLAS:2010mza}
  [ATLAS Collaboration],
  ``Charged particle multiplicities in pp interactions for track $p_{\rm T} > 100$ MeV at $\sqrt{s} =$ 0.9 and 7 TeV measured with the ATLAS detector at the LHC,''
  ATLAS-CONF-2010-046.

\bibitem{Ciesielski:2012mc}
  R.~Ciesielski and K.~Goulianos,
  ``MBR Monte Carlo Simulation in PYTHIA8,''
  PoS ICHEP {\bf 2012} (2013) 301
  [arXiv:1205.1446 [hep-ph]].

\bibitem{Gieseke:2017vli}
  S.~Gieseke, P.~Kirchgaeser and F.~Loshaj,
  ``Soft and diffractive scattering,''
  KA-TP-36-2017, MCnet-17-18
  [arXiv:1710.10925 [hep-ph]].

\bibitem{Biyajima:1984ay}
  M.~Biyajima and N.~Suzuki,
  Prog.\ Theor.\ Phys.\  {\bf 73} (1985) 918;
   Addendum: [Prog.\ Theor.\ Phys.\  {\bf 73} (1985) 1303].

\bibitem{Biyajima:1983qu}
  M.~Biyajima,
  Phys.\ Lett.\  {\bf 137B} (1984) 225; 
   Addendum: [Phys.\ Lett.\  {\bf 140B} (1984) 435].

\bibitem{Fellera:1951aa}
  W. Feller, 
  Ann. Math. {\bf 54} (1951) 173.

\end{thebibliography}
\end{document}